\documentclass[preprint]{aastex7}
\newcommand{\newtext}[1]{{#1}}

\newcommand{\fwidthw}{130mm}  
\newcommand{\fwidthb}{88mm}   
\newcommand{\fwidths}{60mm}   
\usepackage{color}
\usepackage{natbib}
\usepackage{amsmath}
\usepackage{nicefrac}
\tightenlines

\newcommand \Angstrom   {\,{\rm \AA}}

\newcommand \bnhat      {\boldsymbol{\hat{n}}}

\newcommand \brhat      {\boldsymbol{\hat{r}}}

\newcommand \bw         {\boldsymbol{w}}

\newcommand \beq        {\begin{equation}}
\newcommand \beqa	{\begin{eqnarray}}

\newcommand \cm         {\,{\rm cm}}

\newcommand \eeq	{\end{equation}}
\newcommand \eeqa	{\end{eqnarray}}

\newcommand \EM         {{\rm EM}}

\newcommand \erg	{\,{\rm erg}}

\newcommand \eV 	{\,{\rm eV}}

\newcommand \fesc       {f_{\rm esc}}

\newcommand \FWHM       {{\rm FWHM}}
\newcommand \G          {\,{\rm G}}  

\newcommand \gtsim	{\gtrsim}		 

\newcommand \Ha 	{{\rm H}}
\newcommand \He	        {{\rm He}}

\newcommand \K  	{\,{\rm K}}

\newcommand \kms	{\,{\rm km~s}^{-1}}

\newcommand \ltsim	{\lesssim}		 

\newcommand \nH         {n_{\rm H}}

\newcommand \odd        {{\rm o}}

\newcommand \pesc       {p_{\rm esc}}

\newcommand \prand      {{\cal P}}

\newcommand \s	        {\,{\rm s}}
\newcommand \sigmad     {\sigma_{\rm d}}

\newcommand \ttot       {\tau_{\rm tot}}
\newcommand \taud       {\tau_{\rm d}}
\newcommand \taudzero      {\tau_{\rm d,0}}

\newcommand \twotripletP     {$2\,^3{\rm P}^\odd$}
\newcommand \twotripletPzero {$2\,^3{\rm P}_0^\odd$}
\newcommand \twotripletPone {$2\,^3{\rm P}_1^\odd$}
\newcommand \twotripletPtwo {$2\,^3{\rm P}_2^\odd$}
\newcommand \twotripletPJ   {$2\,^3{\rm P}_J^\odd$}
\newcommand \twotripletSone  {$2\,^3{\rm S}_1$}
\newcommand \threetripletSone {$3\,^3{\rm S}_1$}
\newcommand \threetripletP  {$3\,^3{\rm P}^\odd$}
\newcommand \fourtripletP   {$4\,^3{\rm P}^\odd$}

\newcommand{\omittext}[1]{}

\shorttitle{Resonant Scattering of \ion{He}{1} 1.0833$\mu$m Triplet}
\shortauthors{B.T. Draine}
\countdef\decade=200
\advance\decade by \year
\countdef\hours=201
\advance\hours by \time
\divide\hours by 60
\countdef\mins=202
\advance\mins by \hours
\multiply\mins by 60
\multiply\hours by 100
\countdef\miltime=203
\advance\miltime by \hours
\advance\miltime by \time
\advance\miltime by -\mins
\renewcommand\today{\number\decade.\number\month.\number\day.\number\miltime}

\begin{document}

\title{%
        \vspace*{-2.0em}
        {\normalsize\rm {\it The Astrophysical Journal}, accepted}\\ 
        \vspace*{1.0em}
        {\bf Resonant Scattering of the \ion{He}{1} 1.0833$\,\mu$m Triplet in
        \ion{H}{2} Regions: Emission Spectra}
	}

\author[0000-0002-0846-936X]{B.~T.~Draine}
\affiliation{Dept.\ of Astrophysical Sciences,
  Princeton University, Princeton, NJ 08544, USA}
\email{draine@astro.princeton.edu}

\begin{abstract}

Resonant scattering of \ion{He}{1} $1.0833\,\mu$m triplet photons by
metastable He $2\,^3{\rm S}_1$ is studied for optical depths
characteristic of \ion{H}{2} regions.  Regions with large
He\,$2\,^3{\rm S}_1$ column densities are predicted to have unusually
broad multipeaked 1.0833$\,\mu$m emission profiles, with the centroid
blue-shifted by up to $\sim$$14\,{\rm km\,s}^{-1}$ relative to other
lines.  The feature FWHM can exceed $100\,{\rm km\,s}^{-1}$ for some
regions.  \newtext{Resonant trapping enhances dust absorption and
  reduces the \ion{He}{1}\,$1.0833\micron$ emission.  Care must be
  taken when using the
  \ion{He}{1}\,$1.0833\,\mu$m/\ion{H}{1}\,$1.0941\,\mu$m (Pa$\gamma$)
  ratio to estimate the He$^+$/H$^+$ ratio.}  Predicted spectra are
computed for examples, including M17-B and NGC3603 in the Galaxy, and
a star-forming region in M51.  Observations of the $1.0833\,\mu$m
triplet with spectrometers such as NIRSPEC, CARMENES, or X-Shooter can
confirm the predicted effects of resonant scattering in \ion{H}{2}
regions, and constrain the nebular conditions.

\end{abstract}

\section{Introduction}




The \ion{He}{1} $1.0833\micron$ triplet is typically the strongest
observable emission line generated when He$^+$ recombines with
electrons.  It has been observed in emission from \ion{H}{2} regions
powered by O stars \citep{Aller+Liller_1959}, from planetary nebulae
\citep{Liller+Aller_1963,ODell_1963}, from AGNs
\citep{Levan+Puetter+Rudy+etal_1981} and QSOs
\citep{Levan+Puetter+Smith+Rudy_1984}, \newtext{and from the Earth's
  thermosphere \citep[e.g.,][]{Kulkarni_2025}.}

Emission of $1.0833\micron$ triplet photons populates the metastable
He$^0\,1s2s$\,\twotripletSone\ level, and the \twotripletSone\ level
population can become large enough to produce observable absorption.
\citet{Adams_1949} detected interstellar
He$^0$\,\twotripletSone\,--\,\fourtripletP\ $3890\Angstrom$ absorption
toward O stars in the Orion Nebula and the Trifid Nebula, and
absorption in the He$^0$\,\twotripletSone\ --
\twotripletPone\ $1.0833\micron$ triplet itself was observed by
\citet{Galazutdinov+Krelowski_2012} toward the O9.5V star
$\zeta$\,Oph.  In all of these detections, the metastable
He$^0$\,\twotripletSone\ resides in the \ion{H}{2} region around a hot
star.  \citet{Indriolo+Hobbs+Hinkle+McCall_2009} searched for but
failed to detect \ion{He}{1}\,$1.0833\micron$ absorption in the
diffuse neutral ISM outside of \ion{H}{2} regions, placing an upper
limit on the cosmic ray ionization rate.

The 1.0833$\micron$ triplet is also observed in absorption in some
QSOs \citep{Leighly+Dietrich+Barber_2011, Pan+Zhou+Liu+etal_2019},
\newtext{ in the Solar chromosphere
  \citep[e.g.,][]{Pietrow+Kuckein+Verma+etal_2025},} and in transiting
exoplanets with extended atmospheres
\citep{Mansfield+Bean+Oklopcic+etal_2018,
  Nortmann+Palle+Salz+etal_2018, Oklopcic+Hirata_2018,
  Allart+Bourrier+Lovis+etal_2019}.  Spectra of the extragalactic
``Little Red Dots'' also show evidence of $1.0833\micron$ absorption
\citep[e.g.,][]{Wang+deGraaff+Davies+etal_2025}.

\citet{Pottasch_1962} and \citet{Robbins_1968a, Robbins_1968b}
discussed the effects on the populations of other \ion{He}{1} levels
\newtext{in \ion{H}{2} regions and planetary nebulae} when transitions
to \twotripletSone\ become optically thick.  Absorption by the
metastable \twotripletSone\ level can suppress emission in some lines
(e.g., \ion{He}{1}\,$3890\Angstrom$, $3188\Angstrom$) and enhance
emission in certain other lines (e.g., \ion{He}{1}\,$7067\Angstrom$,
$4713\Angstrom$) \citep{Robbins_1968b,Benjamin+Skillman+Smits_2002,
  Blagrave+Martin+Rubin+etal_2007}.
%
%
The theory was applied to interpret \ion{He}{1} line emission
from the Orion Nebula \citep{Blagrave+Martin+Rubin+etal_2007,
  Porter+Ferland+MacAdam_2007}.

Previous studies were concerned with the important effects of resonant
scattering on the overall strengths of \ion{He}{1} emission lines.
The present paper is concerned with the \emph{spectrum} of the
\ion{He}{1}\,1.0833$\micron$ triplet when observed in emission.  For
realistic conditions, the spectrum of the $1.0833\micron$ triplet
photons escaping from the \ion{H}{2} region can be substantially
altered by resonant scattering.  As a result, the $1.0833\micron$
triplet is predicted to have unusual multi-peaked emission profiles.
The \newtext{emission} spectrum is sensitive to the column density of
metastable He$^0$\,\twotripletSone, and to the velocity dispersion of
the recombining He$^+$ that is responsible for populating the
\twotripletSone\ level.  Therefore, observational determination of the
actual line profile will provide a new diagnostic for the nebular
conditions.  It also will serve as a test for our theoretical
understanding of resonant scattering.


The relevant atomic physics is reviewed in Section \ref{sec:1s2s}.
Section \ref{sec:resonant scattering} 
outlines the radiative transfer problem.  Section \ref{sec:dustlessresults}
presents the results of Monte-Carlo simulations
\newtext{for dustless \ion{H}{2} regions, while Section 
5 includes the effects of dust absorption.}  The results are discussed
in Section 
6 and summarized in Section
\ref{sec:summary}.  Details of the Monte Carlo procedures are provided
in the Appendix.

\section{\label{sec:1s2s} The Metastable 2\,$^3{\rm S}_1$ Level}

\begin{figure}
\begin{center}
\includegraphics[angle=0,width=\fwidthw,
                 clip=true,trim=5mm 50mm 0mm 40mm]
{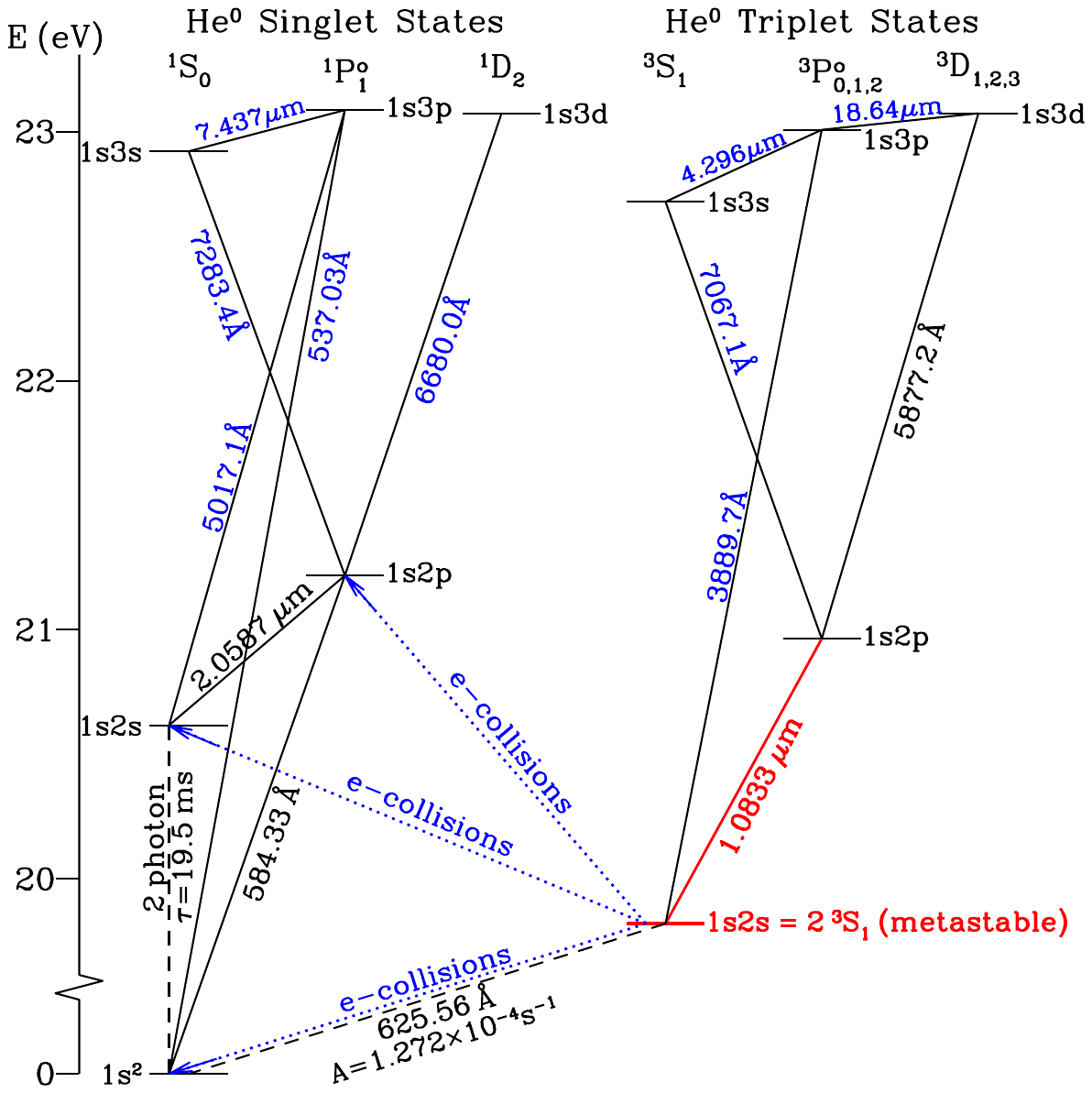}
\caption{\label{fig:He levels} \footnotesize Energy levels of
  \ion{He}{1} below 23.5 eV.  Permitted transitions are shown as solid
  lines, with the $1.0833\micron$ triplet shown in red.  Dashed lines
  indicate forbidden transitions.  Dotted lines show the principal
  paths for collisional depopulation of \twotripletSone\ by
  electrons.
  }
\end{center}
\end{figure}

\subsection{Population by Recombination}

Figure \ref{fig:He levels} shows the first 11 energy levels of \ion{He}{1}.
Approximately 3/4 of the radiative recombinations of He$^+$ with
electrons populate the triplet states of \ion{He}{1}, and essentially
100\% of the triplet recombinations end up populating the lowest energy
triplet state, \twotripletSone.

\omittext{The column density in the metastable level is proportional to the rate
per area of recombination of He$^+$ with electrons to form
triplet \ion{He}{1}:
\beqa
N({\rm He}^0\,2\,^3{\rm S}_1) 
&~=~& 
\frac{ \alpha_{\rm trip}\int n_e n(\He^+)ds }{A_{\rm ms}+n_e k_d}
\\ \label{eq:N(3S1)}
&\approx& 1.90\times10^{14}\cm^{-2} 
\left(\frac{\EM}{10^{24}\cm^{-5}}\right)
\left(\frac{\He^+/\Ha^+}{0.10}\right)
\frac{1}{\left(1+0.26 n_3\right)}
~~~,
\eeqa
where $\EM\equiv \int n_e n(\Ha^+)ds$ is the usual emission measure,}

The $1s2s$ \twotripletSone\ level is metastable, with a
probability per unit time $A_{\rm ms}=1.272\times10^{-4}\s^{-1}$
\citep{Wiese+Fuhr_2009} for decay to the $1s^2\,^1{\rm S}_0$ ground
state with emission of a $625.56\Angstrom$ photon. \newtext{
The density of $\He^0$ \twotripletSone\ is}
\beq
n(2\,^3{\rm S}_1 ) =
\frac{\alpha_{\rm trip}n_e n(\He^+)}{A_{\rm ms} + n_e k_d}
~~~,
\eeq
where \newtext{$n_e$ is the electron density,} $\alpha_{\rm trip}$ is
the rate coefficient for radiative recombination of He$^+$ to the
triplet states of $\He^0$, and $k_d$ is the rate coefficient for
collisional depopulation of \twotripletSone\ by transitions to singlet
states.\footnote{For $T_e=9\times10^3\K$, $\alpha_{\rm
  trip}\approx2.42\times10^{-13}\cm^3\s^{-1}$
\citep{DelZanna+Storey_2022}, and $k_d\approx
3.3\times10^{-8}\cm^3\s^{-1}$ \citep{Bray+Burgess+Fursa+Tully_2000}.
\ion{He}{1}\,\twotripletSone\ can also be depopulated by
photoionization by $h\nu>4.77\eV$ photons, including Lyman-$\alpha$
\citep{Osterbrock_1989}, but this is usually subdominant.}
\begin{figure}
\begin{center}
\includegraphics[angle=0,width=\fwidths,
                 clip=true,trim=0.5cm 5.0cm 0.0cm 4.5cm]
{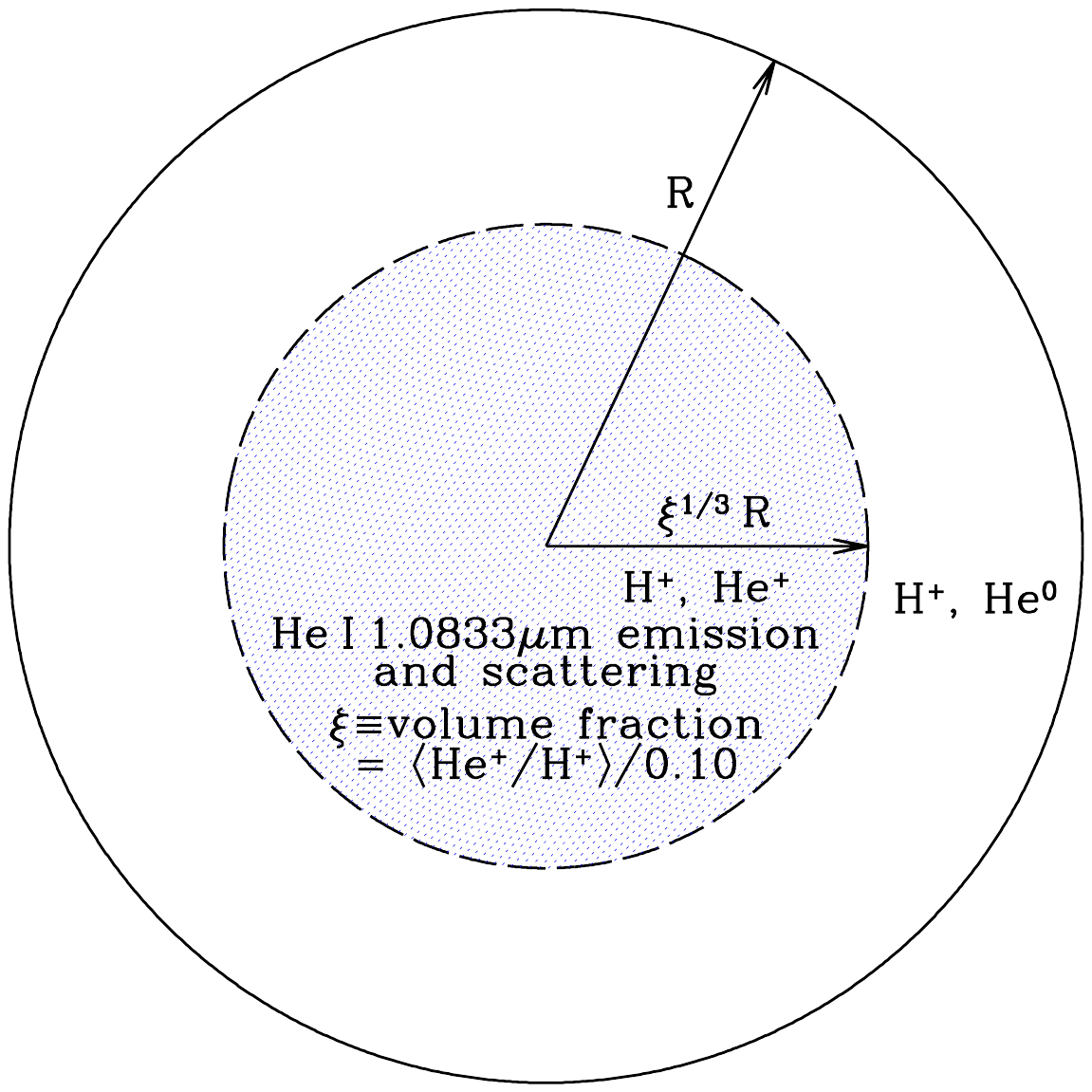}
\caption{\label{fig:sphere}\footnotesize \newtext{\ion{H}{2} region
    cross-section.  $\xi$ is the fraction of the volume where He is
    mainly He$^+$.}  }
\end{center}
\end{figure}

\omittext{Consider a spherical, \newtext{dustless} \ion{H}{2} region
  of radius $R$.  Let $\EM_R$ be the center-to-edge emission measure.}
A uniform density Str\"omgren sphere \omittext{powered by a source
  emitting H-ionizing photons at a}\newtext{with H photoionization}
rate\newtext{\footnote{\newtext{When dust is present, $Q_{48}$ should be
  understood to be the actual H photoionization rate, not the stellar
  output of H-ionizing photons.}}}  $Q_0=10^{48}Q_{48}\s^{-1}$
\newtext{ and $n_e=10^3n_3\cm^{-3}$ has
%
H fully ionized out to radius
\beq
R=9.88\times10^{17}\,Q_{48}^{1/3}\,n_3^{-2/3}\cm
~~~,
\eeq
and center-to-edge emission measure
\beq
\EM_R \equiv \int_0^R n_e \,n(\Ha^+)\,dr = 
9.88\times10^{23}\,Q_{48}^{1/3}\,n_3^{4/3}\cm^{-5}
~~~.
\eeq
 Let $\xi$ be the
  fraction of the volume within which He is singly-ionized (see Figure
  \ref{fig:sphere}).  For helium abundance $n_\He/\nH=0.10$ (as will
  be assumed here), $\xi\approx \min(6.5 Q_1/Q_0,1)$, where $Q_1$ is
  the stellar output of $h\nu>24.6\eV$ photons \citep[see,
    e.g.,][]{Draine_2011a}}
\omittext{center-to-edge H nucleon column density and emission
  measure \beqa \newtext{ N_R({\rm H})} & \newtext{~=~}& \newtext{
  \int_0^R \nH \, dr = \left(\frac{3}{4\pi}\right)^{1/3}
  \frac{Q_0^{1/3} \, n_e^{1/3}}{\alpha_B^{1/3}} } \\ &\newtext{=}&
\newtext{9.88\times10^{20}\, Q_{48}^{1/3}\, n_3^{1/3} \cm^{-2}}
\\ \EM_R &~=~& \int_0^R n_e n(\Ha^+) dr =
\left(\frac{3}{4\pi}\right)^{1/3}
\frac{Q_0^{1/3}\,n_e^{4/3}}{\alpha_B^{1/3}} \\ &=& 9.6\times10^{23}\,
Q_{48}^{1/3}\, n_3^{4/3} \cm^{-5} ~~~.  \eeqa
A single O7V star has $Q_{48}\approx 6$
\citep{Martins+Schaerer+Hillier_2005}.  Thus, an \ion{H}{2} region
with $n_e\approx 10^3\cm^{-3}$, powered by multiple O stars, can have
$\EM_R \gtsim 10^{24}\cm^{-5}$.}  
The center-to-edge column density
\newtext{of metastable He$^0$\,\twotripletSone}
\beq
N_R(2\,^3{\rm S}_1) 
\equiv 
\int_0^R n(2\,^3{\rm S}_1)dr
\approx
1.71\times10^{14}\cm^{-2}
\xi^{1/3}\frac{Q_{48}^{1/3}\,n_3^{4/3}}{(1+0.26n_3)}
\eeq
can exceed $10^{14}\cm^{-2}$.

\begin{table}
\begin{center}
\caption{\label{tab:lambda} \ion{He}{1} 1.0833$\micron$ Triplet$^{a}$}
\begin{tabular}{c c c c c c c c}
\hline
$\ell$                & $u$ & $J_u$ & $g_u$ & $A_{u\ell}\,(\s^{-1})$ & $\lambda\,(\micron)$ &
$\lambda_{\rm air}\,(\micron)$ & $v_u(\kms)\,^b$ \\
\hline
2\,$^3{\rm S}_1$ & 2\,$^3{\rm P}_0^\odd$ & 0   &  3    & $1.0216\times10^7$ & 1.083205748 & 1.082909115
                   & $-29.96$ \\
2\,$^3{\rm S}_1$ & 2\,$^3{\rm P}_1^\odd$ & 1   &  9    & $1.0216\times10^7$ & 1.083321676 & 1.083025011
                   & $+2.19$\\
2\,$^3{\rm S}_1$ & 2\,$^3{\rm P}_2^\odd$ &  2  &  15   & $1.0216\times10^7$ & 1.083330645 & 1.083033978
                   & $+4.67$\\
\hline
2\,$^3{\rm S}_1$ & 2\,$^3{\rm P}^\odd$ & 0,1,2 & 27 & $1.0216\times10^7$ & 1.083313778
                                                   & 1.083017115 & 0\\
\hline
\multicolumn{8}{l}{$^a$ NIST Atomic Physics Database \citep{Kramida+Ralchenko+Reader+NIST_2024}}\\
\multicolumn{8}{l}{$^b$ Doppler shift relative to centroid.}\\
\end{tabular}
\end{center}
\end{table}

\subsection{Radiative Transfer}

Most of the radiative recombinations to the $\He^0$ triplet states
pass through $2\,^3{\rm P}_{0,1,2}^\odd$, resulting in emission of a
photon in the $1.0833\micron$ triplet (see Table \ref{tab:lambda}).
The \twotripletPJ\ created either by direct recombination or by
radiative decay from higher levels will have a Maxwellian velocity
distribution, and newly-injected
\twotripletPJ\,$\rightarrow$\,\twotripletSone\ photons emitted by
these atoms will be created with a Voigt line profile,\footnote{Many
authors approximate the spectrum of the newly-injected photons as a
simple Gaussian profile corresponding to the thermal velocity
distribution, but for large $\tau_0$ the small fraction of photons
injected in the ``damping wings'', with much higher escape
probabilities, should not be neglected.}
characterized by the usual Doppler broadening parameter $b$ and a
``damping constant''
\beq \label{eq:a_ul}
a_{u\ell} \equiv \frac{\gamma_{u\ell}\lambda_{u\ell}}{4\pi b}
~~~,
\eeq
where $\gamma_{u\ell}$ is the sum of the transition probabilities for
depopulating levels $u$ and $\ell$.  Thermal broadening alone gives
\beq
b_{\rm therm} = \left(\frac{2kT}{m_{\He}}\right)^{1/2} = 
6.45 \left(\frac{T}{10^4\K}\right)^{1/2}\kms
~~~;
\eeq
turbulence in the \ion{H}{2} region will result in $b > b_{\rm
  therm}$.  For \ion{H}{2} regions, we consider four cases: $b=6.45$,
$10$, $15$, and $20\kms$.  The \ion{He}{1}\,$1.0833\micron$ triplet
lines each have
\beq
a_{u\ell} \equiv \frac{A_{u\ell}\lambda_{u\ell}}{4\pi b}
= 8.807\times10^{-5}\left(\frac{10\kms}{b}\right)
~~~,
\eeq
\omittext{assuming}\newtext{since} the lifetime of \twotripletSone\ is
long compared to the $10^{-7}\s$ lifetime of \twotripletPJ.

Newly-emitted $1.0833\micron$ triplet photons may be absorbed by
another He atom in the metastable \twotripletSone\ level.  The
\newtext{center-to-edge} line-center optical depth for $2\,^3{\rm S}_1\rightarrow 2\,^3{\rm
  P}_J^\odd$ absorption is
\beqa
\tau_0(2\,^3{\rm S}_1\rightarrow 2\,^3{\rm P}_J^\odd) &~=~& 
(2J+1)
\frac{A_{u\ell}\lambda_{u\ell}^3}{8\pi^{3/2}}
\left(\frac{N_R(2\,^3{\rm S}_1)}{b}\right)
\\
&=& 29.2 (2J+1) 
\left(\frac{N_R(2\,^3{\rm S}_1)}{10^{14}\cm^{-2}}\right)
\left(\frac{10\kms}{b}\right)
~~~.
\eeqa
%

\subsection{\label{sec:multiplet} The $1.0833\mu$m Multiplet}

The three \ion{He}{1} 2\,$^3{\rm S}_1\,-\,2\,^3{\rm P}_J^\odd$ lines
have line shifts
\beq
v_{u\ell} \equiv c \left[1-\frac{\lambda_c}{\lambda_{u\ell}}\right]
\eeq
of $-29.96$, $+2.19$, and $+4.67\kms$ relative to the centroid
wavelength $\lambda_c = 1.0833138\micron$ (see Table
\ref{tab:lambda}).  Because the splittings are small, the multiplet is
often treated as a single line
\citep[e.g.,][]{Robbins_1968b,Benjamin+Skillman+Smits_2002}.  Here we
treat each line separately.

Scattering by a multiplet is identical to scattering by a single line,
except that one must allow for the possibility of the photon being
scattered by any one of the transitions in the multiplet. For a
uniform sphere, the \ion{He}{1}\,$1.0833\micron$ scattering problem is
characterized by two parameters: (1) the Doppler broadening parameter
$b$, and (2) the sum of the line-center optical depths of the three
contributing line profiles,
\beqa \label{eq:ttot}
\ttot &~\equiv~&
\sum_{J=0}^2 \tau_0(2\,^3{\rm S}_1\rightarrow\,2\,^3{\rm P}_J^\odd) 
= 262 \left(\frac{N_R(2\,^3{\rm S}_1)}{10^{14}\cm^{-2}}\right)
\left(\frac{10\kms}{b}\right)
\\ \label{eq:ttot_vs_Q48n3}
&\newtext{\approx}&
\newtext{448 ~
\frac{\xi^{1/3} Q_{48}^{1/3}n_3^{4/3}}{(1+0.26n_3)}
\left(\frac{10\kms}{b}\right)
}
~~~.
\eeqa

\section{\label{sec:resonant scattering} Resonant Scattering}

Resonant scattering for a single line was discussed by
\citet{Unno_1952}, \citet{Field_1959b}, \citet{Hummer_1962}, and many
subsequent papers, with particular attention to the scattering of the
H Lyman $\alpha$ line
\citep[e.g.,][]{Bonilha+Ferch+Salpeter+etal_1979,Neufeld_1990,
  Zheng+Miralda-Escude_2002,Dijkstra_2019,Seon+Kim_2020}.  Here we
consider resonant scattering of \ion{He}{1}\,$1.0833\micron$
triplet photons.

We assume a spherical \ion{H}{2} region with uniform electron density
$n_e$ and He$^+$ density $n(\He^+)$.  The recombining He$^+$ results
in a uniform density of metastable \ion{He}{1}\,\twotripletSone.
$1.0833\micron$ triplet photons are randomly injected with initial
frequency drawn from the Voigt profile for emission from
\twotripletPJ\ with the Maxwellian velocity distribution of the
recombining He$^+$.  Once emitted, the photons travel until they
either scatter or escape the \ion{H}{2} region. The scattering is
treated using the full 3-line multiplet, with a Voigt absorption
profile for each line.
 
If scattering takes place, the scattered photon travels in a new
direction with a new frequency.  The new direction is drawn from the
scattering phase function for electric dipole scattering.  The new
frequency is drawn from the ``partial redistribution function''
$R_{\rm II-B}$ \citep{Hummer_1962} that describes coherent scattering
by a Maxwellian distribution of atoms with finite natural
linewidth.\footnote{The frequency redistribution function is sampled
using the ``rejection method''
\citep{Press+Teukolsky+Vetterling+Flannery_1992,Zheng+Miralda-Escude_2002}.}
\omittext{Dust is neglected, and Thomson scattering by electrons is
  assumed to be negligible.} \newtext{The present radiative transfer
  treatment neglects Thomson scattering by electrons, but allows for
  absorption by dust.}  Our Monte-Carlo methodology for scattering in
a multiplet is described in the Appendix.  Complete numerical results
can be obtained from \url{https://doi.org/10.7910/DVN/EOHGDD}.

\section{\label{sec:dustlessresults}\omittext{Results}\newtext{\ion{He}{1}$\,1.0833\micron$ Emission from Dustless \ion{H}{2} Regions}}

\subsection{Number of Scatterings and Escape Probability}

\begin{figure}
\begin{center}
\includegraphics[angle=0,width=\fwidthb, 
                 clip=true,trim=0.5cm 5.0cm 0.0cm 4.5cm]
{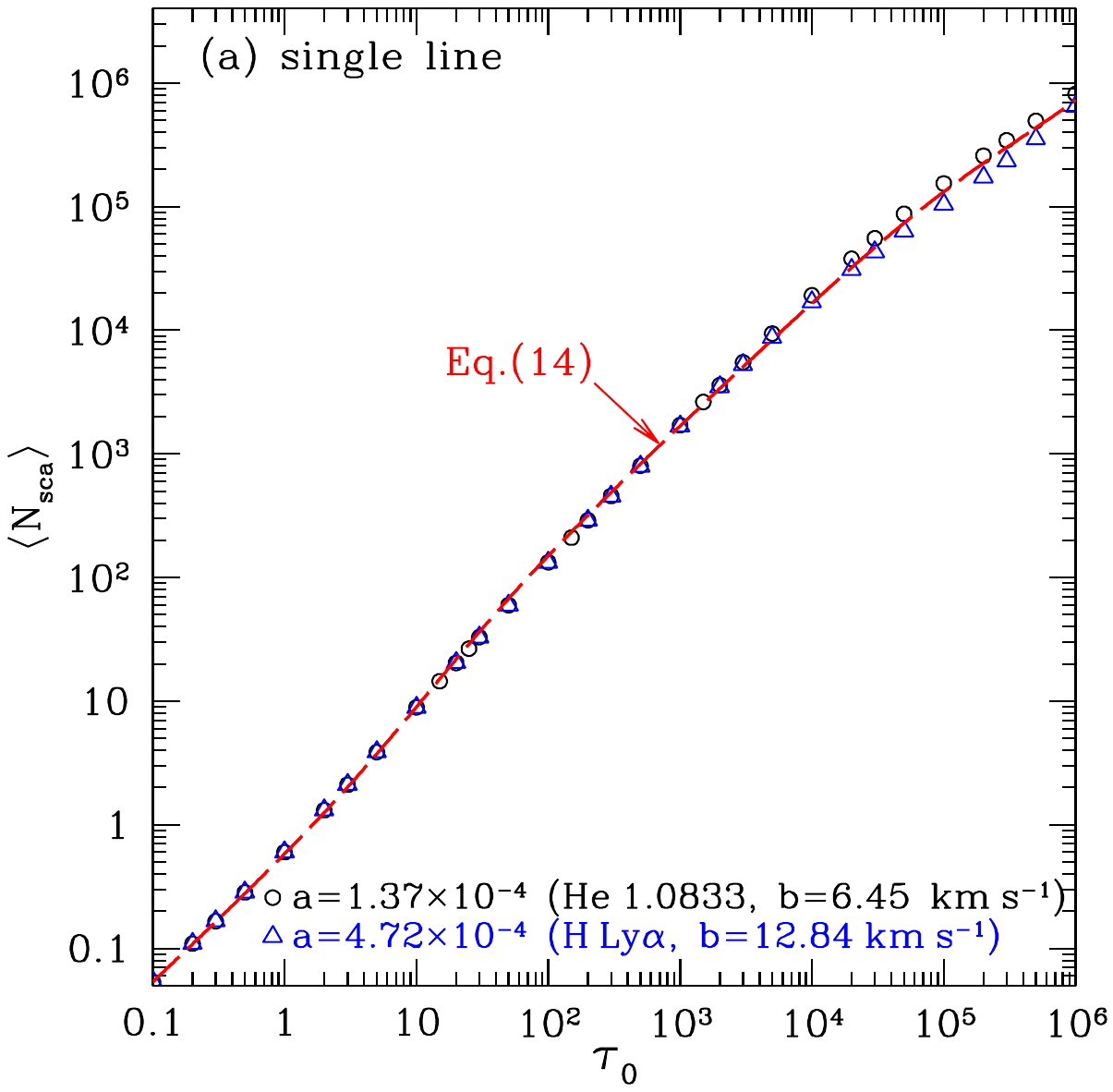}
\includegraphics[angle=0,width=\fwidthb,
                 clip=true,trim=0.5cm 5.0cm 0.0cm 4.5cm]
{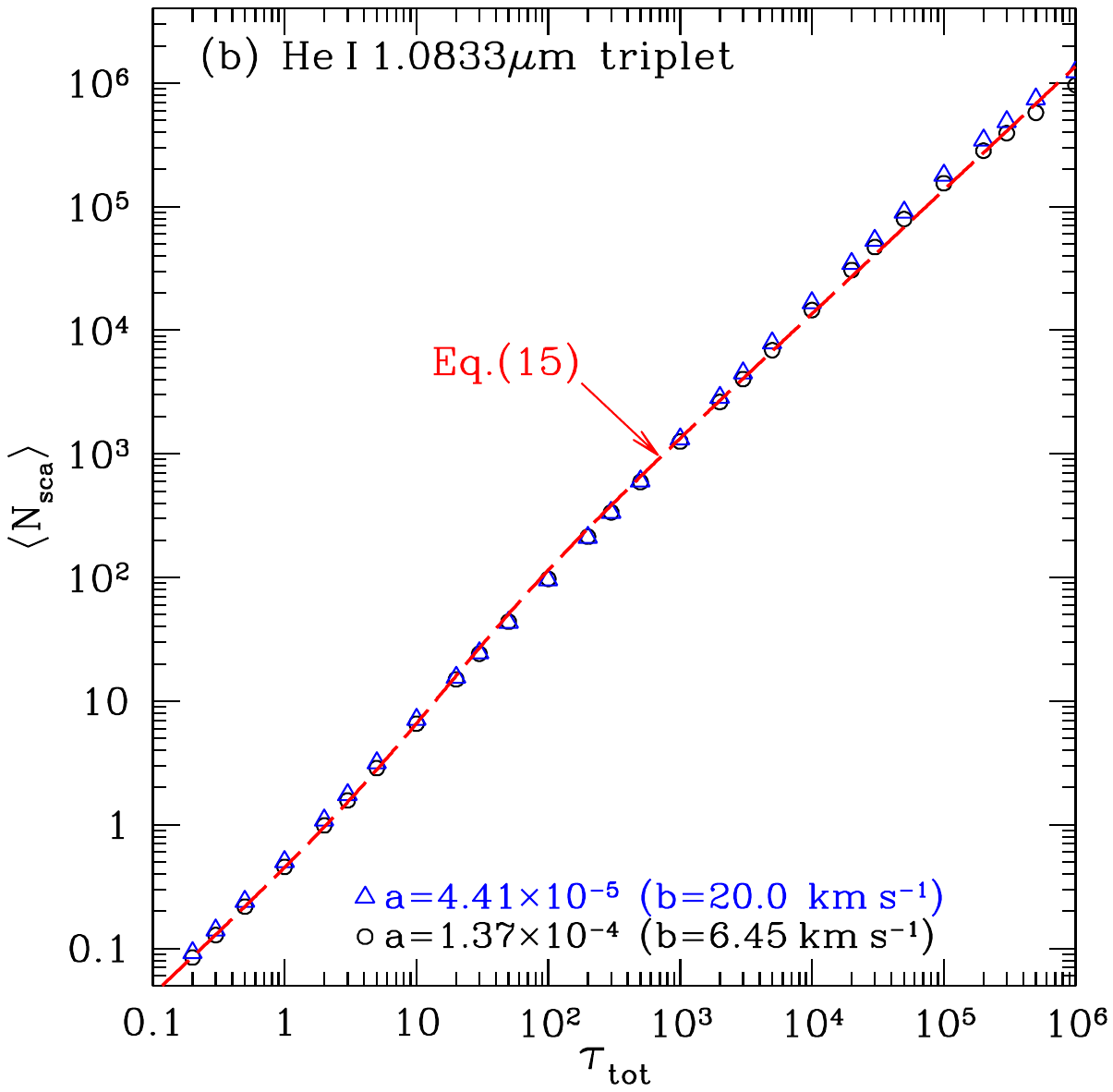}
\caption{\label{fig:nsca}\footnotesize Number of scatterings for
  spherical geometry for (a) single resonance line 
  \newtext{(e.g., Lyman\,$\alpha$)}, for two values of
  the damping constant $a$.  (b) \ion{He}{1}\,$1.0833\micron$ triplet,
  for two values of $b$.  
  \newtext{The approximations (\ref{eq:nsca1})
  and (\ref{eq:nsca3}) are shown as dashed red curves.}
}
\end{center}
\end{figure}

\newtext{Let $\tau_0$ be the line-center optical depth from center to
  edge of a spherical \ion{H}{2} region.}  \omittext{The present
  study} \newtext{If no dust is present, there are no}\omittext{does
  not include any} processes leading to photon loss, hence every
injected triplet photon eventually escapes.  Let $\langle N_{\rm
  sca}\rangle$ be the mean number of scatterings per injected photon.
We define the ``escape probability'' to be
\beq \label{eq:beta}
\beta \equiv \frac{1}{1+\langle N_{\rm sca}\rangle}
~~~,
\eeq
i.e., the fraction of the escape attempts that succeed.  

For scattering by a single resonant line \newtext{(e.g., Lyman
  $\alpha$, or \ion{He}{1}$\,1.0833\micron$ approximated as a single
  line)}, $\langle N_{\rm sca}\rangle$ is shown as a function of
$\tau_0$ in Figure \ref{fig:nsca}a, for two values of the damping
constant $a$.  For given $\tau_0$, $\langle N_{\rm sca}\rangle$ is
almost independent of $a$ -- note the small difference between
$\langle N_{\rm sca}\rangle$ for $a=1.37\times10^{-4}$ and
$4.72\times10^{-4}$.

For scattering by a single line (e.g., Lyman\,$\alpha$), $\langle
N_{\rm sca}\rangle$\omittext{ and $\beta$} can be approximated by the fitting
function
\beq \label{eq:nsca1}
N_{\rm sca,1} \approx
\frac{3}{4\sqrt{2}}\tau_0 
\left[ 1 + \frac{0.1\tau_0}{(1+0.045\tau_0)(1+5\times10^{-6}\tau_0)}\right]
~~~,
\eeq
shown in Figure \ref{fig:nsca}a.  Equation (\ref{eq:nsca1}) is exact
in the limit $\tau_0\rightarrow 0$, and provides a good approximation
for single-line scattering for $\tau_0 \ltsim 10^6$.
%

\begin{figure}
\begin{center}
\includegraphics[angle=0,width=\fwidthb,
                 clip=true,trim=0.5cm 5.0cm 0.0cm 4.5cm]
{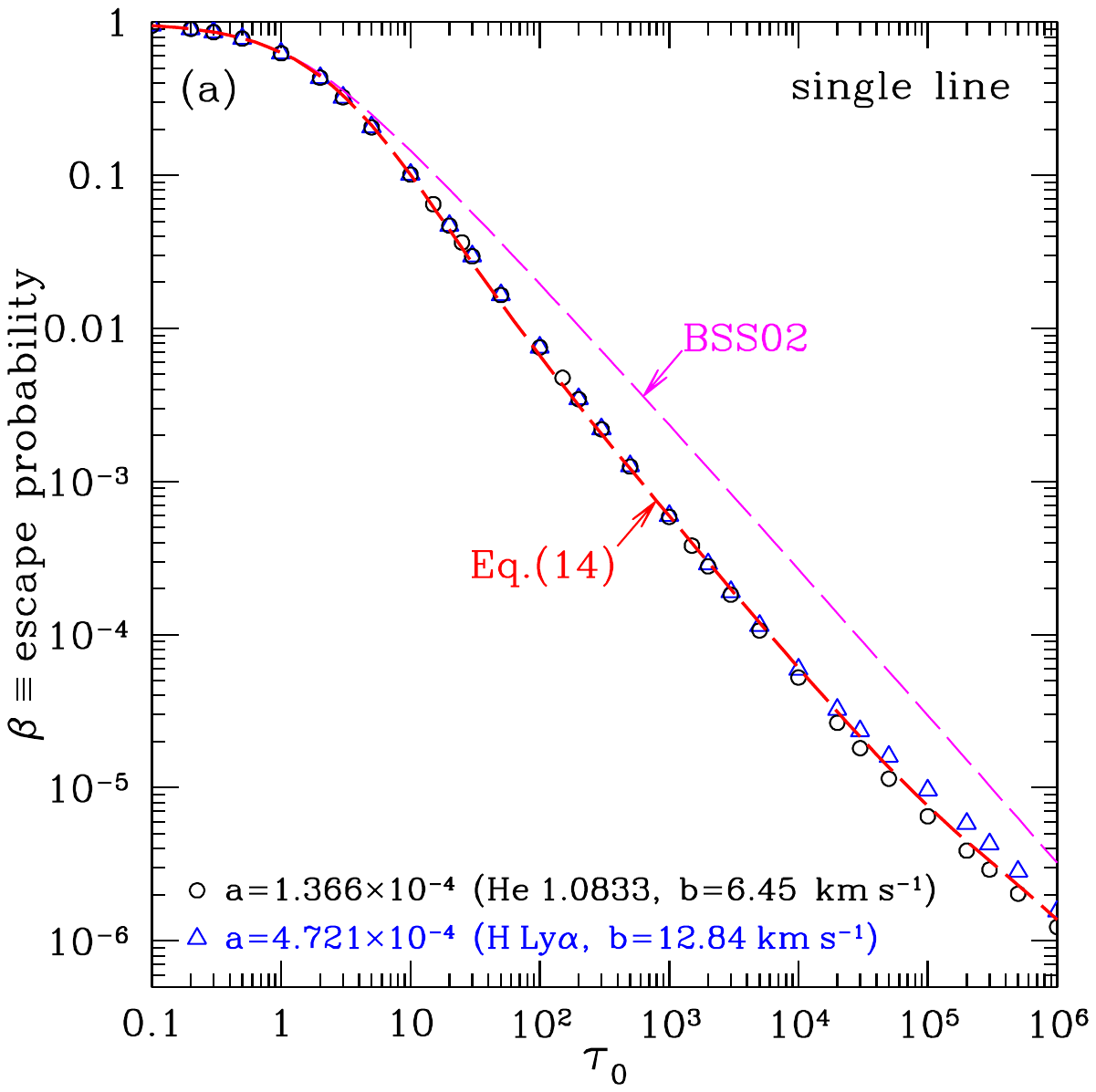}
\includegraphics[angle=0,width=\fwidthb,
                 clip=true,trim=0.5cm 5.0cm 0.0cm 4.5cm]
{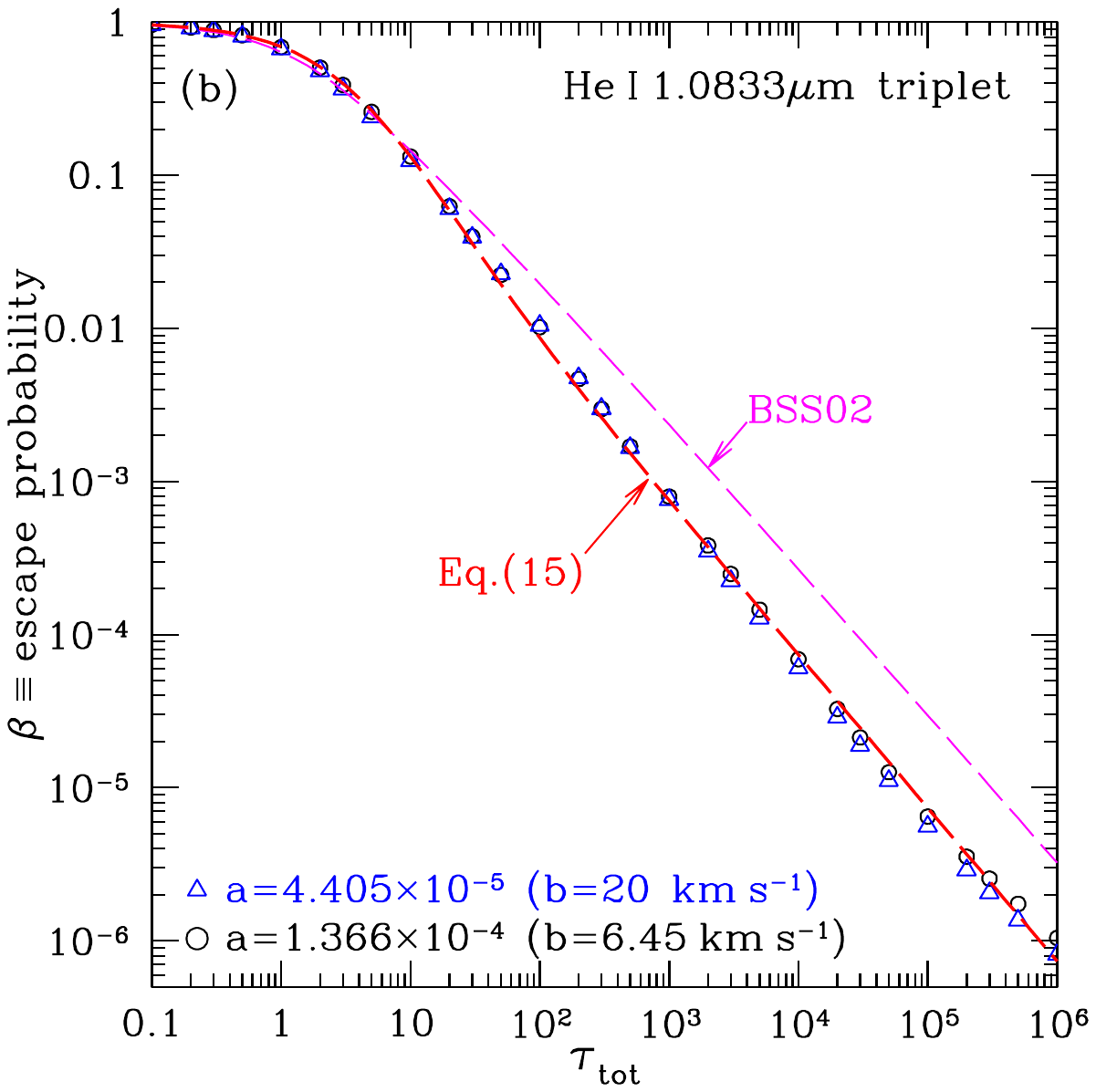}

\caption{\label{fig:beta}\footnotesize Escape probability $\beta\equiv
  1/(1+\langle N_s\rangle)$ as a function of optical depth. (a) Single
  resonant line, for spherical geometry, and $a=1.37\times10^{-4}$
  (appropriate for \ion{He}{1}\,1.0833$\micron$) and
  $a=4.72\times10^{-4}$ (appropriate for \ion{H}{1}\,Ly$\alpha$).
  Also shown: $\beta$ calculated using Eq.\,(\ref{eq:nsca1}), and the
  escape probability $\beta_{\rm BSS02}$ estimated by
  \citet{Benjamin+Skillman+Smits_2002}.  (b) Full three-line treatment
  for the \ion{He}{1} $1.0833\micron$ triplet for spherical geometry,
  for two values of $b$.  Also shown: $\beta$ calculated using
  Eq.\ (\ref{eq:nsca3}), and $\beta_{\rm BSS02}$.
  }
\end{center}
\end{figure}

Figure \ref{fig:nsca}b shows $\langle N_{\rm sca}\rangle$ calculated
for the \ion{He}{1}\,$1.0833\micron$ triplet, for two values of $b$.
As for the single-line case in Figure \ref{fig:nsca}a, $\langle N_{\rm
  sca}\rangle$ is almost independent of $b$. While $\langle N_{\rm
  sca}\rangle$ is similar to the single-line result, it differs in
detail.  For the \ion{He}{1} triplet, $\langle N_{\rm
  sca}\rangle$\omittext{ and $\beta$} can be approximated by
\beq \label{eq:nsca3}
N_{\rm sca,3} \approx 0.42\ttot
\frac{(1+0.11\ttot)}{(1+0.034\ttot)}
~~~,
\eeq
shown in Figure \ref{fig:nsca}b.  Equation (\ref{eq:nsca3}) provides a
good approximation for $\langle N_{\rm sca}\rangle$ for the
\ion{He}{1}\,$1.0833\micron$ triplet for $\ttot\ltsim 10^6$, for a
range of $b$ values.

\begin{figure}
\begin{center}
\includegraphics[angle=0,width=\fwidthb,
                 clip=true,trim=0.5cm 5.0cm 0.0cm 4.5cm]
{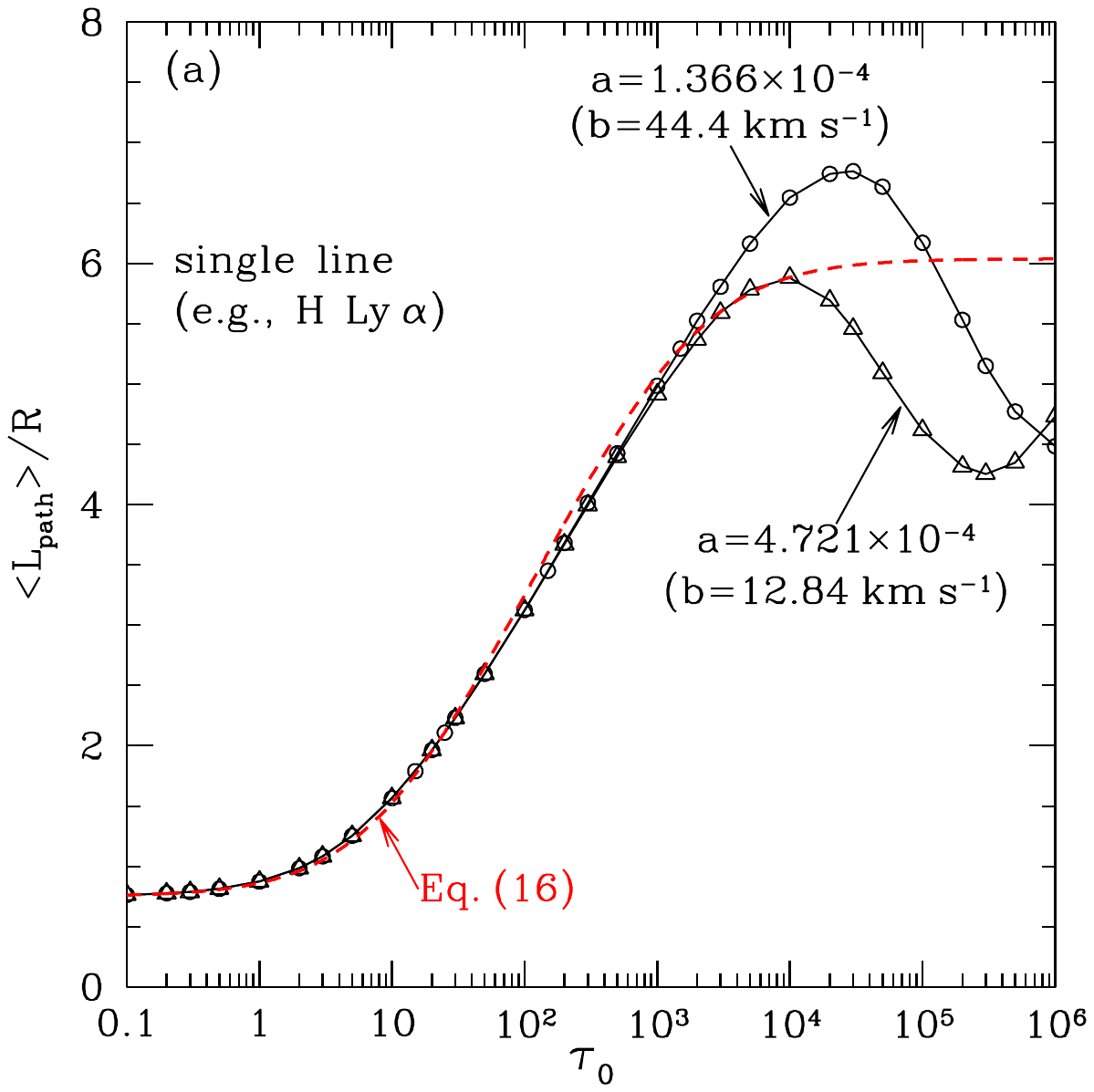}
\includegraphics[angle=0,width=\fwidthb,
                 clip=true,trim=0.5cm 5.0cm 0.0cm 4.5cm]
{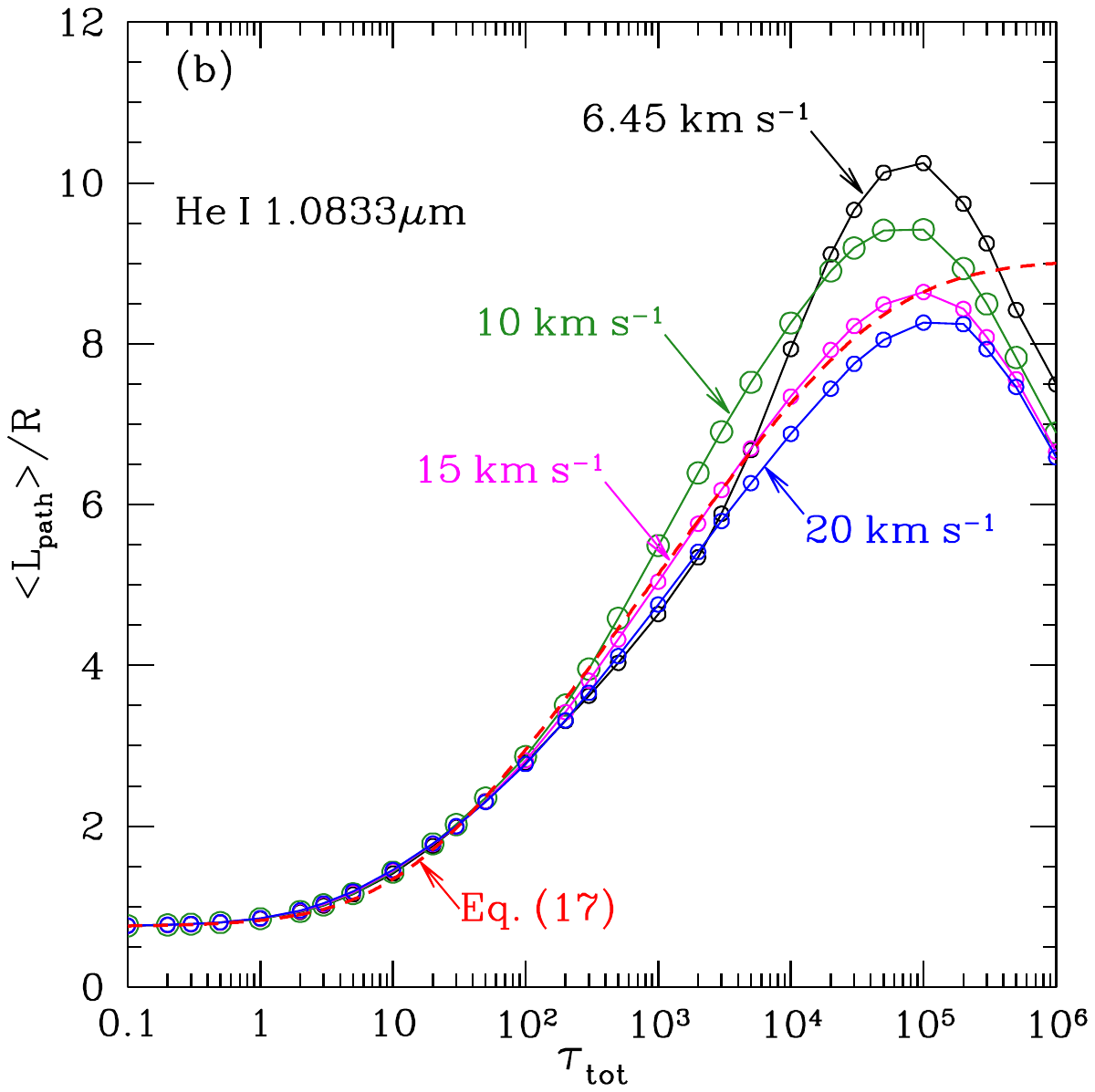}
\caption{\footnotesize\label{fig:path} $\langle L_{\rm
    path}\rangle/R$, where $\langle L_{\rm path}\rangle$ is average
  path length traveled by photons before escaping from a sphere of
  radius $R$.  (a) Results for a single line (e.g., Ly$\alpha$) for
  two values of $a$.  The dashed line is an empirical fit
  (Eq.\ \ref{eq:Lfita}).  (b) Results for the \ion{He}{1}
  $1.0833\micron$ triplet for 4 values of $b$.  The dashed line is an
  empirical fit (Eq.\ \ref{eq:Lfit}).
  }
\end{center}
\end{figure}

The escape probability $\beta$ defined by Equation (\ref{eq:beta}) is
shown for single-line scattering in Figure \ref{fig:beta}a.
\citet[][hereafter BSS02]{Benjamin+Skillman+Smits_2002} approximated
the triplet as a single line, and estimated the escape probability by
integrating the ``escape function'' $\epsilon(\tau_\nu)$
\citep{Cox+Mathews_1969} over a Maxwellian line profile.  This
integral, labelled $\beta_{\rm BSS02}$, is shown in Figure
\ref{fig:beta}.  For scattering of a single line in a spherical
nebula, $\beta_{\rm BSS02}$ overestimates $\beta$ for $\ttot\gtsim
10$, exceeding $\beta$ by a factor $\sim$$5$ at $\ttot=10^4$.

It is not surprising that our calculation of $\beta$ differs from
$\beta_{\rm BSS02}$: they are different quantities.  Our definition of
$\beta$ (see Equation \ref{eq:beta}) is based on the mean number of
scatterings per injected photon, allowing the spectrum to be modified
by the scattering process, whereas $\beta_{\rm BSS02}$ is defined to
be the probability of escape \emph{without} scattering for photons
with a Maxwellian line profile.

Figure \ref{fig:beta}b shows $\beta$ calculated for the
\ion{He}{1}\,$1.0833\micron$ triplet, for $b=6.45$ and $20\kms$.  For a
single line, $\beta(\tau_0)$ is independent of $b$, except for a weak
dependence via the damping constant $a$.  For the full 3-line
treatment for the \ion{He}{1}\,$1.0833\micron$ triplet, $\beta(\ttot)$
depends only weakly on $b$.  Comparing Figures \ref{fig:beta}a and
\ref{fig:beta}b, we see that $\beta$ calculated using the full
three-line treatment for \ion{He}{1}\,$1.0833\micron$ is similar to --
but slightly below -- the results of the single-line approximation.
For $6\ltsim b \ltsim 20\kms$, the fitting function (\ref{eq:nsca3})
provides a good approximation for scattering of \ion{He}{1}
triplet photons (see Figure \ref{fig:beta}b).

\subsection{Path Length}

Scattering affects the path length traversed by photons before
escaping the scattering volume.  Figure \ref{fig:path} shows the mean
distance $\langle L_{\rm path}\rangle$ traveled by photons before
leaving the sphere.  As $\ttot$ increases, $L_{\rm path}/R$ increases
slowly, peaking near $\ttot\approx 10^5$.  Figure \ref{fig:path}a
shows results for scattering by a single line (e.g., Ly\,$\alpha$),
for two different values of $a$.  For $\tau\ltsim 10^4$, $\langle
L_{\rm path}\rangle$ is approximated by
\beq \label{eq:Lfita}
\frac{\langle L_{\rm path}\rangle}{R}\approx\ln\left[e^{3/4}+\frac{0.25\tau_0}{(1+0.0006\tau_0)}\right]
\hspace*{10mm}{\rm for~a~single~line}
~~~.
\eeq
The path length for\omittext{ scattering of } \ion{He}{1}
1.0833$\micron$ triplet photons is shown in Figure \ref{fig:path}b for
$\ttot<10^6$.  For $\ttot\ltsim 10^{4.5}$, $\langle L_{\rm
  path}\rangle$ can be approximated by the fitting function
\beq \label{eq:Lfit}
\frac{\langle L_{\rm path}\rangle}{R} ~\approx~
\ln\left[e^{3/4}+\frac{0.17\ttot}{(1+2\times10^{-5}\ttot)}\right]
\hspace*{10mm}{\rm for~He\,I}~1.0833\micron
~~~.
\eeq
%

\begin{figure}
\begin{center}
\includegraphics[angle=0,width=\fwidthb,
                 clip=true,trim=0.5cm 5.0cm 0.5cm 4.5cm]
{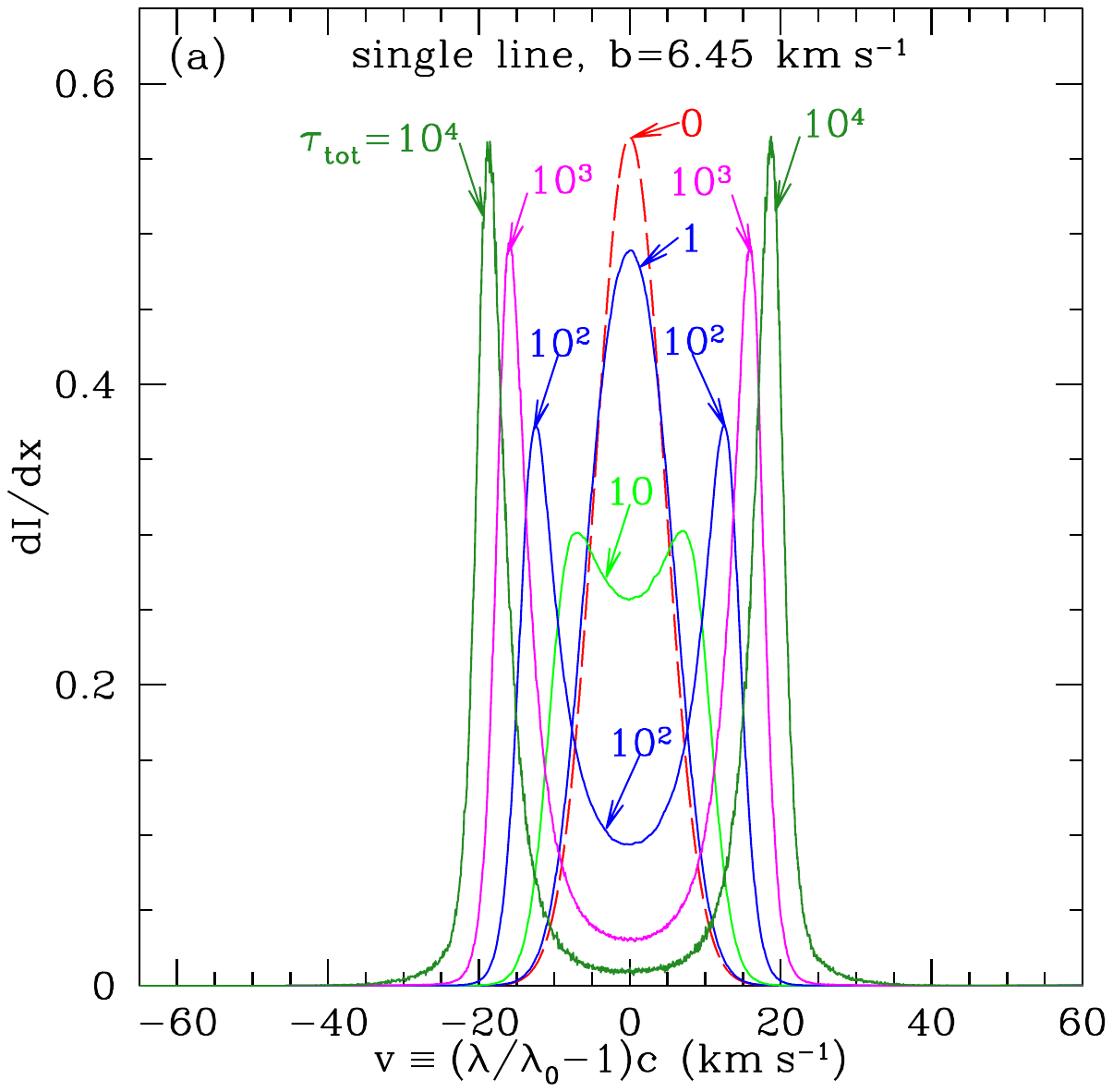}
\includegraphics[angle=0,width=\fwidthb,
                 clip=true,trim=0.5cm 5.0cm 0.5cm 4.5cm]
{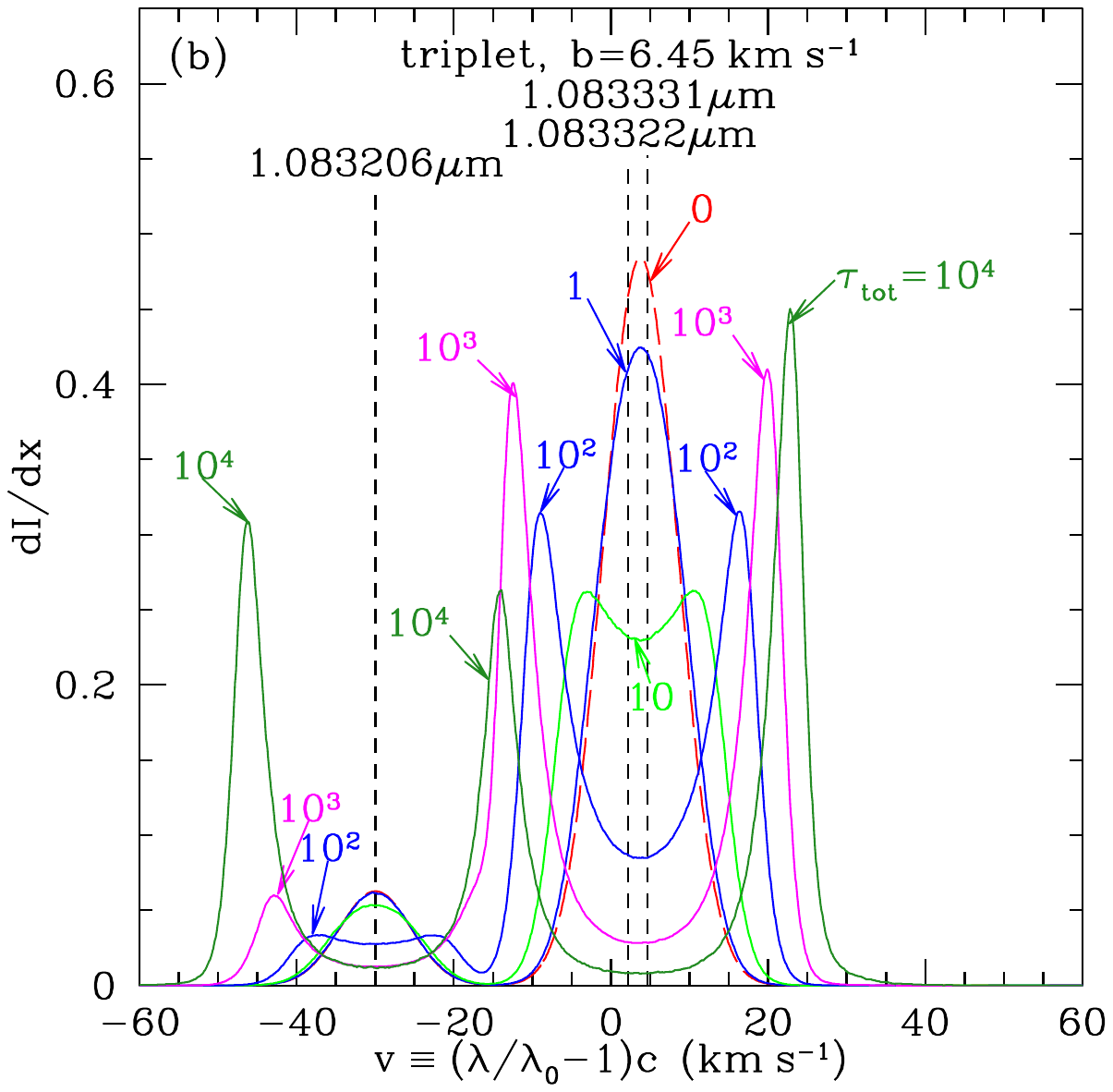}
\caption{\label{fig:spec}\footnotesize Spectra $dI/dx$ of escaping
  photons for \newtext{dustless \ion{H}{2} regions with} $b=6.45\kms$,
  for various values of the line-center optical depth $\ttot$.  The
  normalization is $\int I dx=1$, where
  $x\equiv(\lambda_0/\lambda-1)(c/b)$. \newtext{Velocity $v$ is
    relative to the centroid $\lambda_0$ for optically-thin
    absorption.} (a) Single-line scattering \newtext{(e.g.,
    Lyman\,$\alpha$)}. (b) The \ion{He}{1}\,1.0833$\micron$ triplet.
  Spectra are shown for $\ttot=0, 1, 10, 10^2, 10^3$, and $10^4$;
  profiles are labelled by $\ttot$.  The single-line profiles (panel
  a) are symmetric, while the triplet profiles (panel b) are
  asymmetric.  For the same $b$ and $\ttot$, the triplet profile is
  blue-shifted and broader than would be predicted for a single-line
  treatment.
  }
\end{center}
\end{figure}

The residence time within the \ion{H}{2} region per injected photon is
just $L_{\rm path}/c$, hence the energy density of $1.0833\micron$
triplet photons will be
\beq
\langle u_{1.0833}\rangle = 
\frac{\alpha_{\rm trip}n_e n({\rm He}^+) R}{c}
\left(\frac{\langle L_{\rm path}\rangle}{R}\right)
\left(\frac{hc}{\lambda}\right)
~~~.
\eeq
%
%

\medskip

\subsection{Spectrum}

\begin{figure}
\begin{center}
\includegraphics[angle=0,width=\fwidthb,
                 clip=true,trim=0.5cm 5.0cm 0.5cm 4.5cm]
{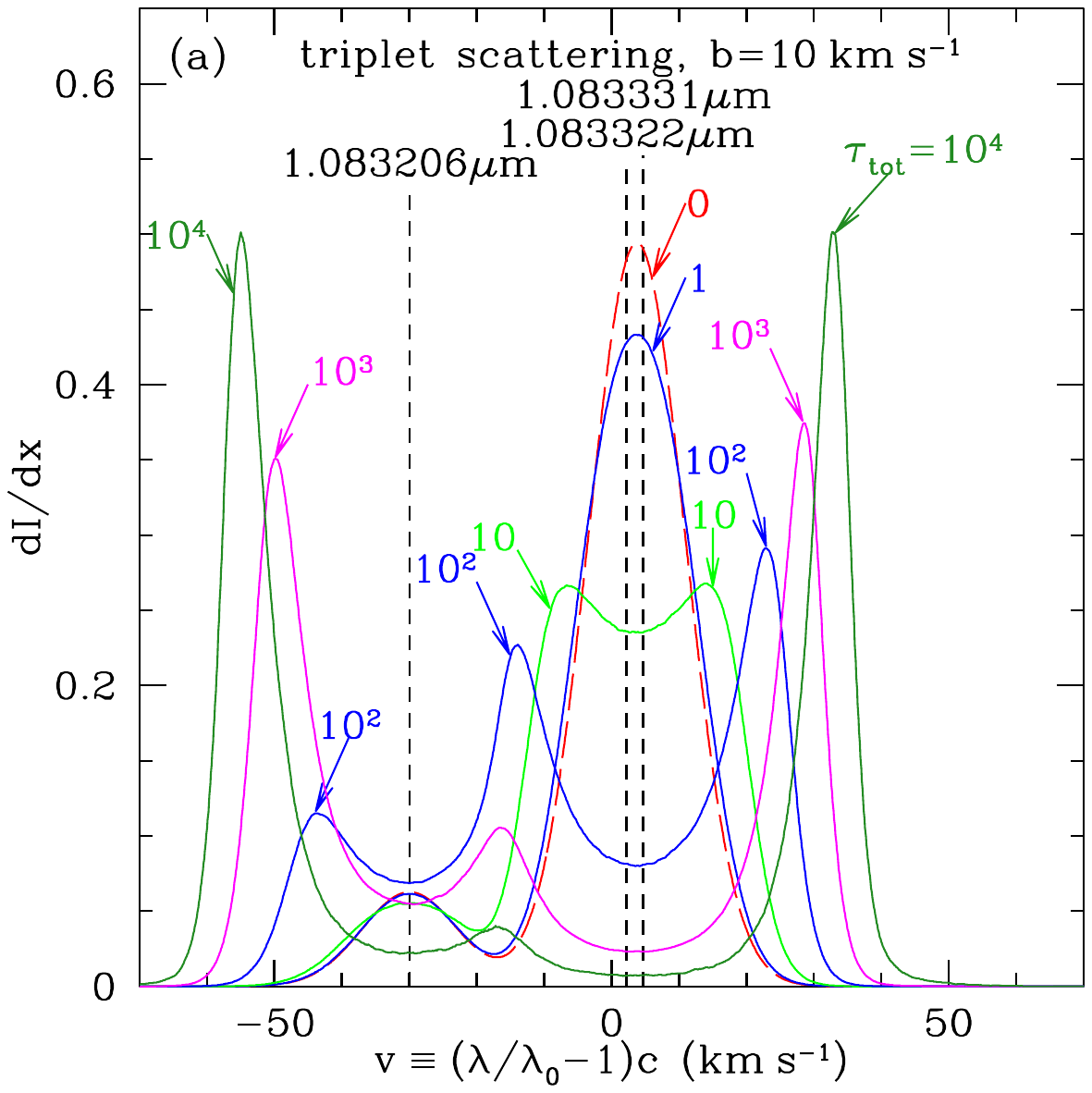}
\includegraphics[angle=0,width=\fwidthb,
                 clip=true,trim=0.5cm 5.0cm 0.5cm 4.5cm]
{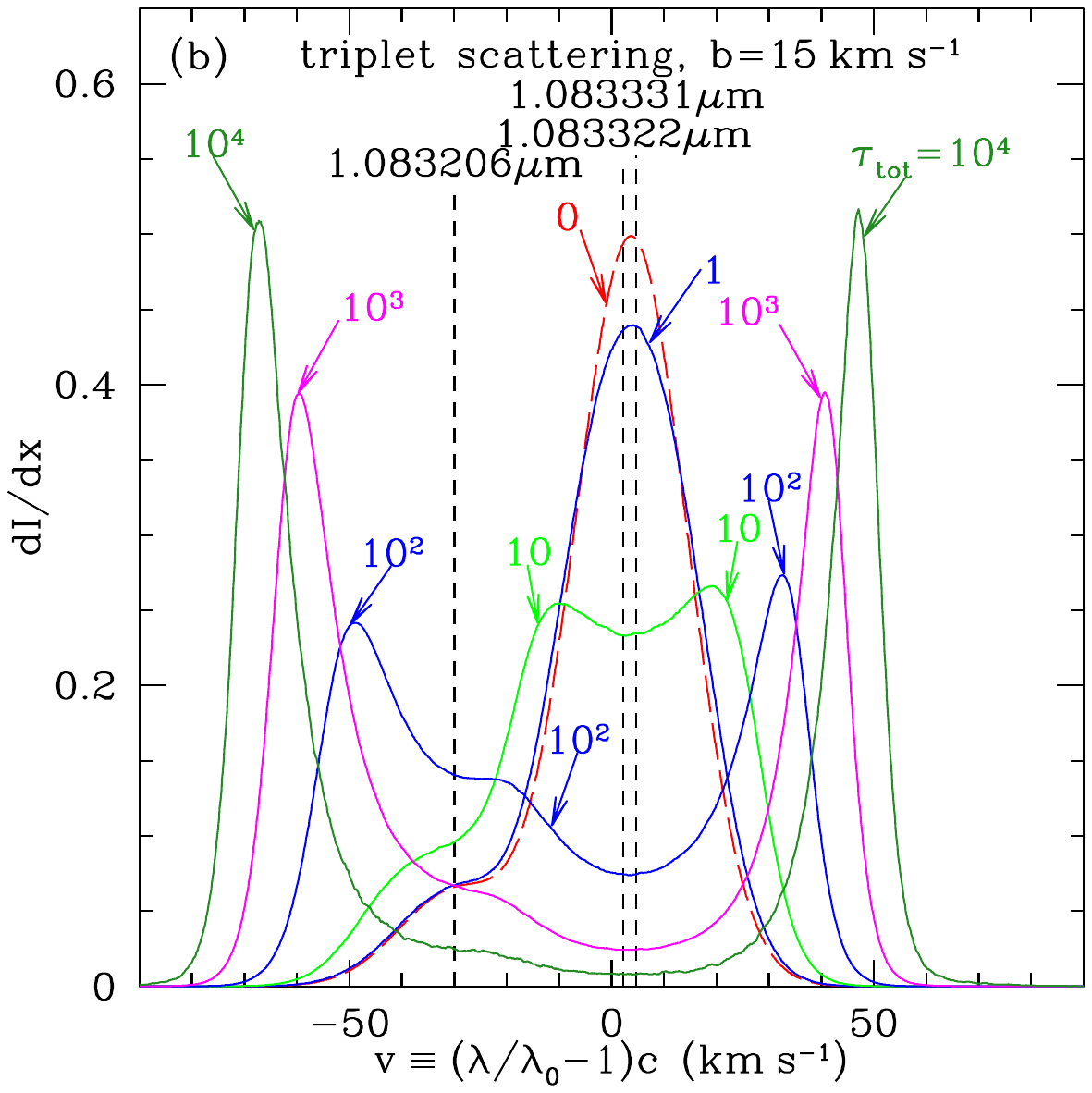}
\caption{\label{fig:specb}\footnotesize Same as Figure
  \ref{fig:spec}b, but for (a) $b=10\kms$, and (b) $b=15\kms$.  For
  each case, results are shown for $\ttot=0, 1, 10, 10^2, 10^3$, and
  $10^4$; curves are labelled by $\ttot$.  Note the increased
  splitting for $b=15\kms$ relative to $b=10 \kms$ (panel a) or
  $b=6.45\kms$ (Figure \ref{fig:spec}b).
  }
\end{center}
\end{figure}

For single-line scattering, the spectrum of the
escaping photons is shown in Figure \ref{fig:spec}a for $b=6.45\kms$
and five values of $\ttot$.  Single-line scattering generates a
symmetric line profile.  For $\ttot \ltsim 5$, the profile is
centrally-peaked, while for $\ttot\gtsim 10$ it develops a central
minimum, with two symmetric peaks; as $\ttot$ increases, the central
minimum deepens, and the peaks are increasingly red-shifted and
blue-shifted.  Figure \ref{fig:spec}a shows the profiles for
$b=6.45\kms$, but the problem is self-similar,\footnote{Except for a
weak dependence on $a\propto b^{-1}$.}
with the profile width (for fixed $\tau_0$) proportional to $b$.

Figure \ref{fig:spec}b shows line profiles calculated for the same
velocity dispersion as in Figure \ref{fig:spec}a, and the same values
of $\ttot$, but for the full $1.0833\micron$ triplet.  The line
profiles in Figure \ref{fig:spec}b are qualitatively different from
Figure \ref{fig:spec}a.

The \emph{shape} of the profile of escaping \ion{He}{1}
$1.0833\micron$ photons depends on $b$ as well as $\ttot$.  Our
discussion here follows the results for $b=6.45\kms$ (Figure
\ref{fig:spec}b).  In the optically-thin limit $\ttot\ltsim 1$, the
line profile resolves into two components; 8/9 of the power is in the
blended emission at $1.08333\micron$ from \twotripletPone\ and
\twotripletPtwo, and 1/9 of the power is in the $1.08321\micron$ line
from \twotripletPzero.  As $\ttot$ increases, the line profile departs
from the simple optically-thin profile, as shown in Figures
\ref{fig:spec}b, \ref{fig:specb}a, and \ref{fig:specb}b.  First, the
blended emission from \twotripletPone\ and \twotripletPtwo\ broadens,
and for $\ttot\gtsim5$ develops a central minimum, so that for
$\ttot=10$ the line profile has three peaks: two of the peaks are
approximately symmetric around the (blended) \twotripletPone\ and
\twotripletPtwo\ lines, and the third is at the wavelength of the
weaker emission at $1.08321\micron$ from \twotripletPzero.  For
$\ttot\gtsim50$ the $1.08321\micron$ line also develops a central
minimum.  For $\ttot \gtsim 10^4$ the outer two peaks become
increasingly dominant, moving farther apart in wavelength with
increasing $\ttot$.

\begin{figure}
\begin{center}
\includegraphics[angle=0,width=8.0cm,
                 clip=true,trim=0.5cm 5.0cm 0.0cm 4.5cm]
{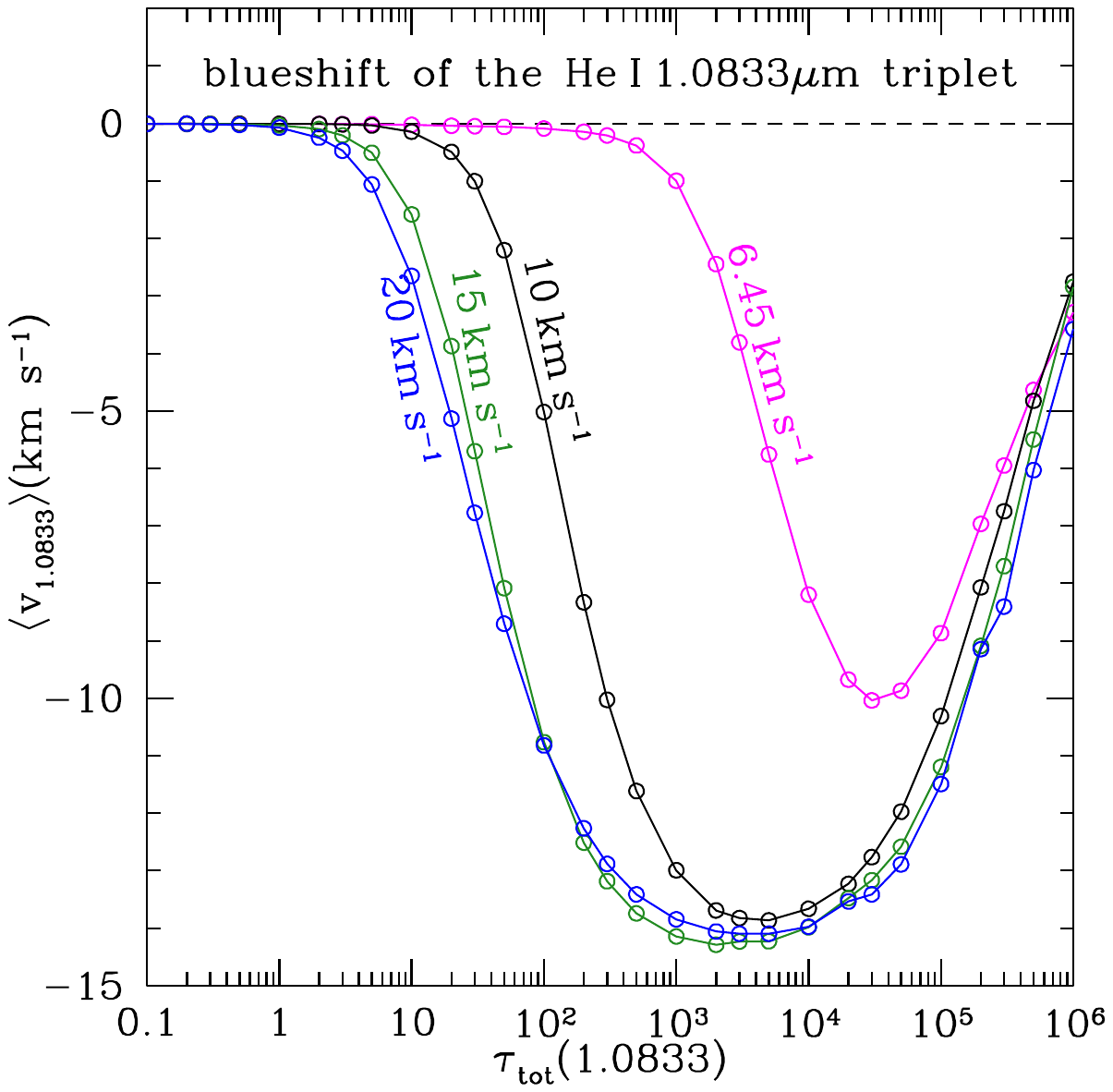}
\caption{\label{fig:blueshift}\label{fig:linewidth}\footnotesize
  Overall blue shift of radiation from the \ion{He}{1} $1.0833\micron$
  triplet, as a function of the optical depth $\tau_{\rm tot}$ of the
  \ion{He}{1} $1.0833\micron$ triplet, for \ion{He}{1} velocity
  $b=6.45,10,15,20\kms$.
  }
\end{center}
\end{figure}
\begin{figure}
\begin{center}
\includegraphics[angle=0,width=\fwidthb,
                 clip=true,trim=0.5cm 5.0cm 0.0cm 4.5cm]
{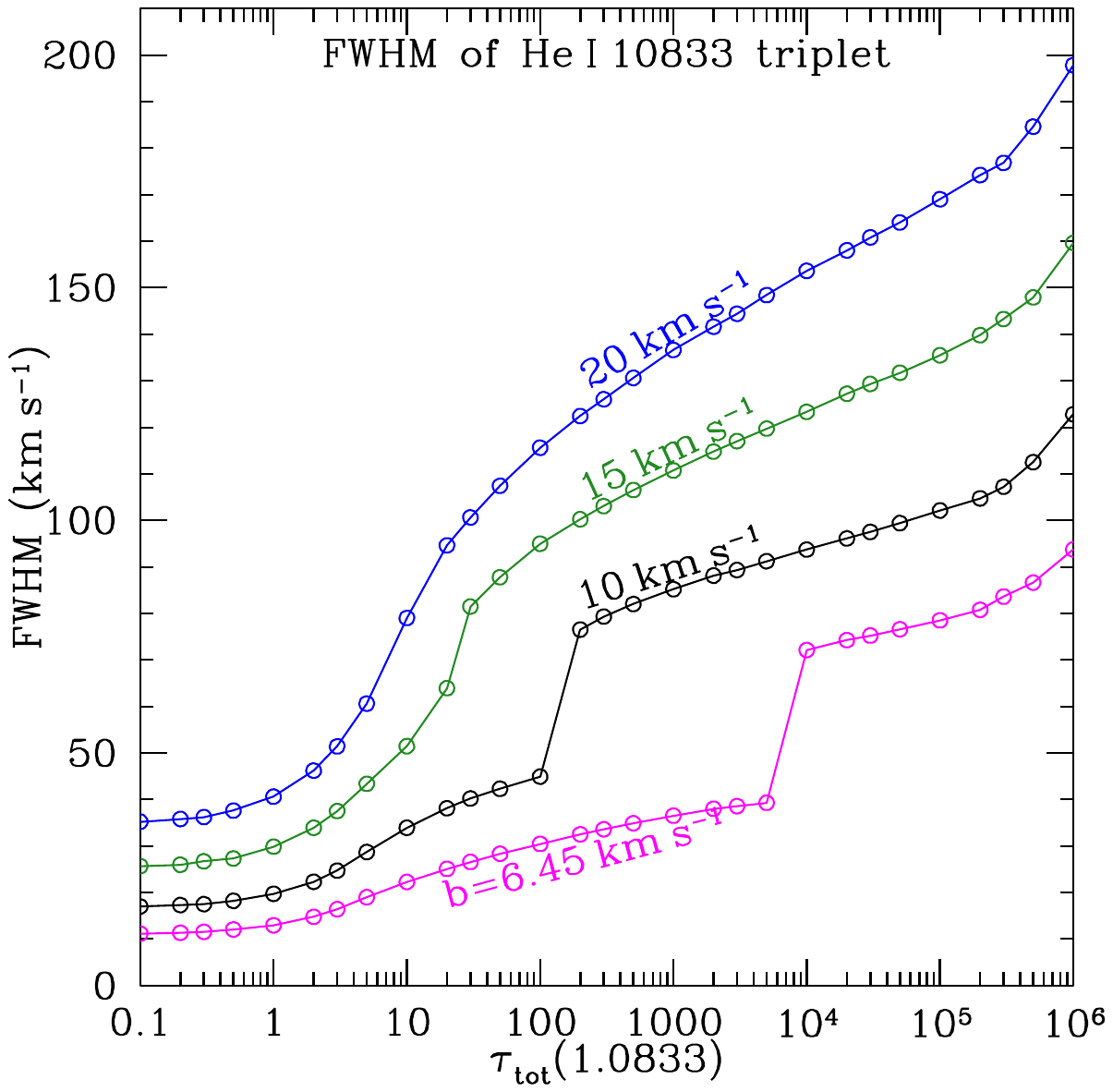}
\caption{\label{fig:linewidth}\footnotesize 
  Full width at half maximum (FWHM) of the line profile (in velocity
  units), as a function of the optical depth parameter $\ttot$, for
  four values of the Doppler broadening parameter $b$.
  }
\end{center}
\end{figure}
\begin{figure}
\begin{center}
\includegraphics[angle=0,width=\fwidthb,
                 clip=true,trim=0.5cm 5.0cm 0.0cm 4.5cm]
{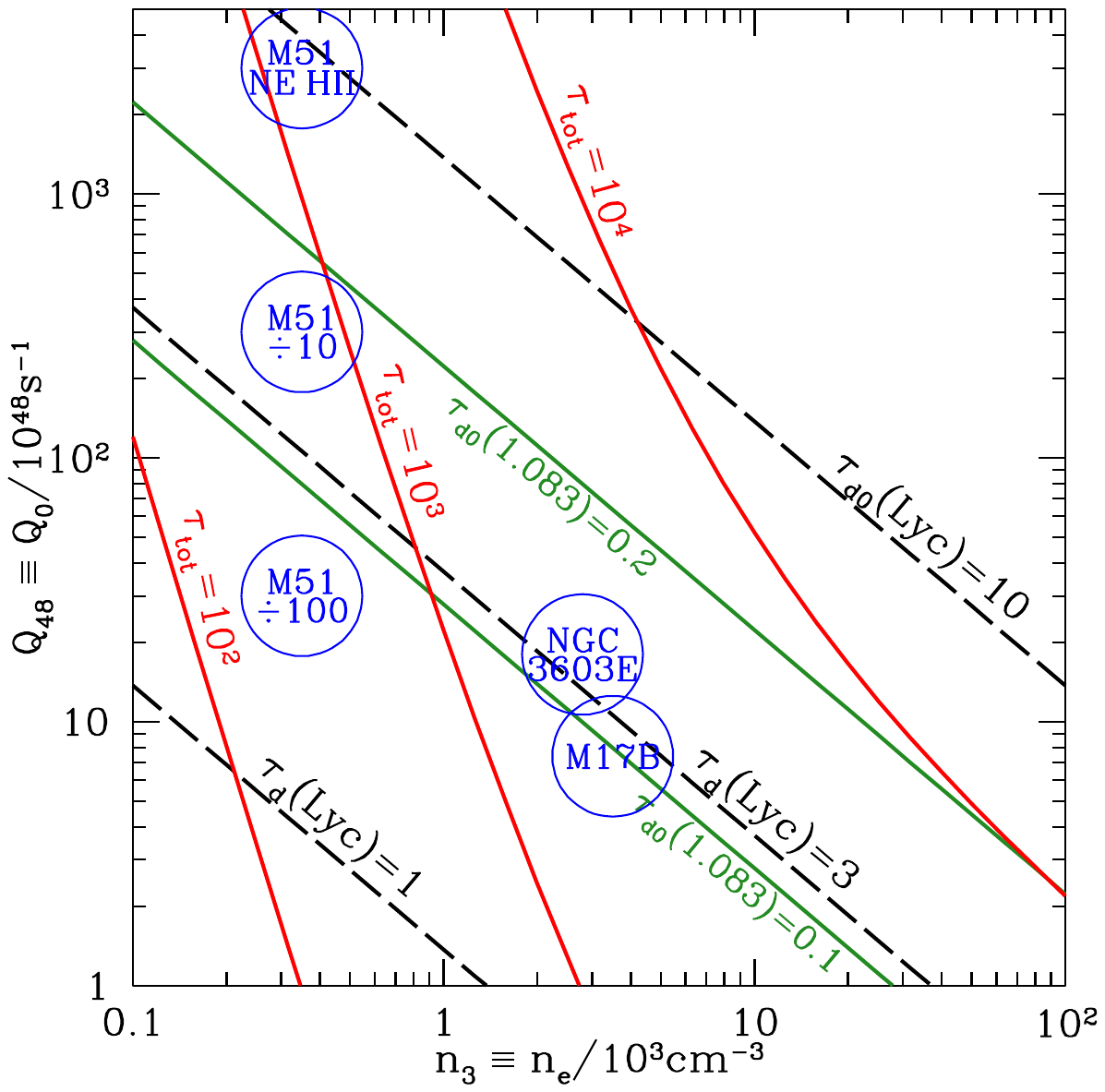}
\caption{\label{fig:parspace}\footnotesize \newtext{ Red lines:
    contours of constant $1.0833\micron$ scattering optical depth
    $\ttot$ (for $\xi=1$) on the $n_3-Q_{48}$ plane.  Black lines:
    contours of constant dust optical depth $\taudzero$ for Lyman
    continuum (dashed) and $1.0833\micron$ (green).  Also shown are
    $(n_3,Q_{48})$ for M17B, NGC\,3603\,Source E, and three
    possibilities for NE-Strip\,\ion{H}{2} in M51 (see Section
    \ref{sec:discuss}).}  }
\end{center}
\end{figure}

The one-line treatment generates a symmetric profile, with no overall
redshift or blue shift.  The three-line treatment, however, produces
an overall blue shift of the emitted photons when $\ttot\gtsim 10$,
because some of the photons emitted by $2\,^3{\rm P}_{1,2}^\odd$
scatter blueward and are then subject to scattering in the $2\,^3{\rm
  S}_1 - 2\,^3{\rm P}_0^\odd$ transition.  The mean blueshift of the
escaping photons is shown in Figure \ref{fig:blueshift}.  For $b\gtsim
10\kms$ and $100\ltsim\ttot\ltsim 3\times10^4$, Figure
\ref{fig:blueshift} shows that the \ion{He}{1} $1.0833\micron$ triplet
is predicted to have a mean blueshift exceeding $10\kms$ relative to
the other \ion{He}{1} and \ion{H}{1} recombination lines.

In addition to producing a systematic blue shift of the \ion{He}{1}
$1.0833\micron$ emission, resonant scattering broadens the line
profile.  Figure \ref{fig:linewidth} shows the
full-width-at-half-maximum (FWHM) of the emission profile.  For small
$\ttot$ the FWHM has values appropriate to optically-thin emission:
for small $\ttot$ the FWHM is not affected by the weak line because it
remains below 50\% of the peak.  As $\ttot$ is increased, the FWHM
increases discontinuously when the line profile develops a peak
blueward of $1.0832\micron$ that rises to 50\% of the peak intensity.
For $b\approx 10\kms$, the FWHM can exceed $75\kms$ for $\ttot\gtsim
10^2$.

\newtext{

\section{\label{sec:dustyresults} \ion{He}{1}\,$1.0833\micron$ Emission from
Dusty \ion{H}{2} Regions}

\subsection{Parameter Space}
}

%
%

\newtext{Figure \ref{fig:parspace} shows (in red) contours of
  constant $\ttot$ on the $n_3-Q_{48}$ plane, for dustless Str\"omgren
  spheres, and assuming He$^+$/H$^+=0.10$.  The locations of M17B and
  NGC\,3603\,Source E are indicated.  Three possibilities are shown for
  NE-Strip\,\ion{H}{2} in M51.  It is evident that many \ion{H}{2}
  regions are expected to have $\ttot\gtsim 10^3$, with strong
  resonant scattering of the \ion{He}{1}\,$1.0833\micron$ triplet.


The optical properties of dust within \ion{H}{2} regions are
uncertain.  For dust absorption cross section $\sigmad({\rm
  Lyc})\approx1\times10^{-21}\cm^2$ and
$\sigmad(1.0833\micron)\approx 3.5\times10^{-23}\cm^2$ per H nucleon
\citep{Hensley+Draine_2020,Gordon+Clayton+Decleir+etal_2023}\footnote{%
\newtext{We take the absorption cross section $\sigmad=3.5\times10^{-23}\cm^2/\Ha$ from the Astrodust model
\citep{Hensley+Draine_2023}.}}, 
the center-to-edge absorption optical depths for the \ion{H}{2} region
are
\beqa
\taudzero({\rm Lyc})&~\approx~& 0.9 \, Q_{48}^{1/3}\,n_3^{1/3}
\\ \label{eq:taud_vs_Q48n3}
\taudzero(1.0833\micron)&\approx& 0.033 \, Q_{48}^{1/3}\,n_3^{1/3}
~~~.
\eeqa
and the dust optical depth from the center to the edge of the
\ion{He}{2} zone is
\beq
\taud(1.0833\micron) ~=~ \xi^{1/3}\taudzero
~~~.
\eeq
Figure \ref{fig:parspace} also shows contours of constant $\taud({\rm
  Lyc})$ (in black) and $\taudzero(1.0833\micron)$ (in green) on the
$n_3-Q_{48}$ plane.  For example: an \ion{H}{2} region with
$Q_{48}=10^2$, $n_3=1$, $\xi=1$, and $b=10\kms$ would have
$\ttot\approx1800$ (from Equation \ref{eq:ttot_vs_Q48n3}), and
center-to-edge $\taud(1.0833\micron)\approx 0.16$ (from Equation
\ref{eq:taud_vs_Q48n3}).  }

%

\newtext{For center-to-edge optical depth $\taud$ for dust absorption,
  nonresonant lines (such as \ion{H}{1}\,Pa$\gamma$) are suppressed by a
  factor $\fesc(\taudzero)$, where
\beqa \label{eq:F(taud)}
\fesc(x) &~\equiv~& \frac{3}{8 x^3}
\left[2 x^2 - 1 + \left( 1+2 x \right) e^{-2 x}\right]
\\
&=& 1 - \frac{3}{4}x + \frac{2}{5}x^2 -\frac{1}{6}x^3 ... ~~~~~{\rm for~}x\ll 1
\eeqa
is the exact result for the photon escape fraction from a sphere with
uniform emission per volume and uniform absorption with center-to-edge
optical depth $x$.

With resonant trapping enhancing the path length traveled, the escape
probability for \ion{He}{1}\,$1.0833\micron$ triplet photons is
reduced.  A simple estimate for the escape probability is
\beq \label{eq:approx pesc}
\pesc ~\approx~ 
\fesc\left(\taud\times\frac{\langle L_{\rm path}\rangle}{R}\right)
~~~,
\eeq
where $\langle L_{\rm path}\rangle/R$ is given by Eq.\ (\ref{eq:Lfit})
(based on calculations for the dustless case).


\begin{figure}
\begin{center}
\includegraphics[angle=0,width=\fwidthb,
                 clip=true,trim=0.5cm 5.0cm 0.0cm 4.5cm]
{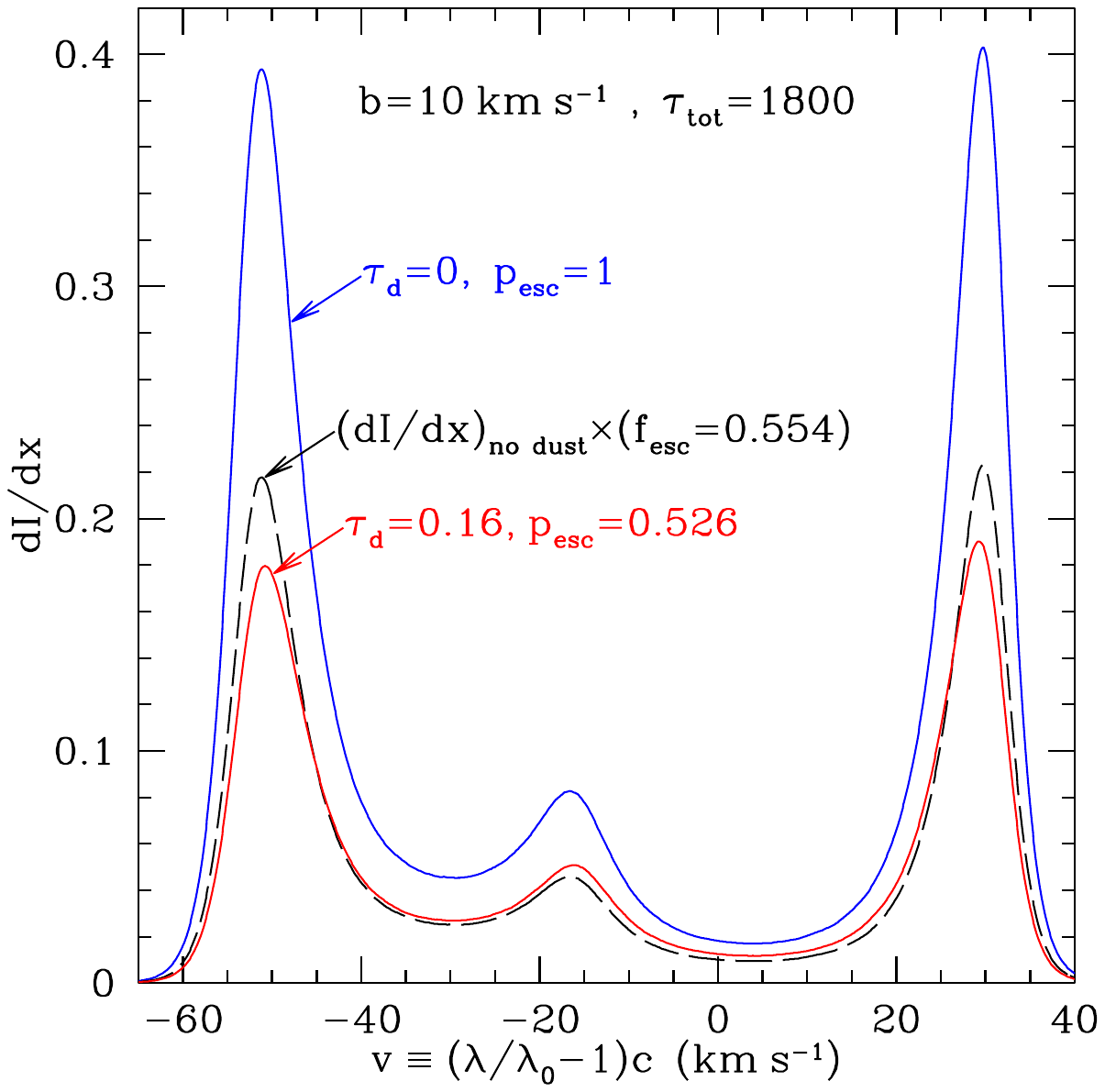}
\caption{\label{fig:dust}\footnotesize \newtext{For $b=10\kms$ and
    $\ttot=1800$: Spectrum of the escaping photons in the absence of
    dust (blue), and for dust with center-to-edge absorption optical
    depth $\taud(1.0833\micron)=0.16$ (red curve).  The black broken
    curve is the dustless spectrum multiplied by $f_{\rm
      esc}(\taud\times\langle L\rangle/R)$, with $\langle L\rangle/R$
    obtained from Equation (\ref{eq:Lfit}).  For this case, the
    approximation slightly overestimates the line power by $\sim$5\%:
    $f_{\rm esc}=0.554 > p_{\rm esc}=0.526$).  The shape of the
    spectrum with dust is similar to the attenuated dustless spectrum,
    but differs in detail: lower at the red-shifted and blue-shifted
    peaks, but slightly higher at the central peak).}}
\end{center}
\end{figure}

As a test, Figure \ref{fig:dust} shows the results of accurate
Monte-Carlo calculations for $\ttot=1800$, $b=10\kms$, for both
$\taud=0$ (blue curve) and $\taud=0.16$ (red curve).  For this case,
Equation (\ref{eq:Lfit}) predicts $\langle L_{\rm path}\rangle/R
\approx 5.70$, and Equation (\ref{eq:approx pesc}) gives $p_{\rm
  esc}\approx f_{\rm esc}(0.911) = 0.554$, in good agreement with
$p_{\rm esc}= 0.526$ from the Monte-Carlo calculation, confirming that
Equation (\ref{eq:approx pesc}) provides a reasonable estimate for
$p_{\rm esc}$ if direct Monte-Carlo calculations are unavailable.  The
spectrum for $\taud=0$, scaled down by the estimate $f_{\rm
  esc}=0.554$ (broken curve), is in fair agreement with the actual
spectrum for $\taud=0.16$ (red curve).  This indicates that the suite of
spectra calculated for dustless \ion{H}{2} regions can be used (with
Equations \ref{eq:approx pesc} and \ref{eq:Lfit}) to estimate
\ion{He}{1}\,$1.0833\micron$ for dusty \ion{H}{2} regions for
$\taud\ltsim0.2$.

For $Q_{48}n_3 \gtsim 10$ (i.e., $\taudzero({\rm Lyc})\gtsim 2$),
absorption of ionizing radiation by dust becomes significant, and
radiation pressure on dust and gas causes the \ion{H}{2} to develop a
shell-like structure \citep{Draine_2011c}.  For $Q_{48}n_3 \gtsim 10$,
accurate calculations of the emission spectrum should allow for the
nonuniform density of both dust and gas.  However, in the absence of
such calculations, a reasonable approximation is to use Equations
(\ref{eq:approx pesc}) and (\ref{eq:Lfit}) to estimate the effects of
dust, with $Q_{48}$ in Equation (\ref{eq:ttot_vs_Q48n3}) denoting the
actual H photoionization rate.  

Real \ion{H}{2} regions with $Q_{48}\gtsim 10^2$ will be powered by
many O stars, and the distribution of both the ionizing stars and the
gas may be complex.  In addition, there may be velocity gradients
arising from gas flows.  It will not be surprising if actual spectra
deviate from the static, spherically-symmetric models considered here.
%
}

\medskip

\newtext{
\subsection{The \ion{He}{1}\,$1.0833\micron$/\ion{H}{1}\,Pa\,$\gamma$ 
            Line Ratio}

The
\ion{He}{1}\,$1.0833\micron$/\ion{H}{1}\,Pa\,$\gamma\,1.09441\micron$
line ratio is useful for measuring the He$^+$/H$^+$ ratio, because the
nearly identical wavelengths imply that differential extinction due to
foreground dust will not affect the line ratio.  However, the line
ratio will be affected by radiative trapping of the
\ion{He}{1}\,$1.0833\micron$ photons within the \ion{H}{2} region.

\ion{He}{1}\,$1.0833\micron$ is both emitted and resonantly trapped
only in the region where He is ionized; if $\xi < 1$, this will be
only the central part of the \ion{H}{2} region.  The
\ion{He}{1}\,$1.0833\micron$/\ion{H}{1}\,Pa\,$\gamma$ ratio for
radiation emerging from the \ion{H}{2} region can be estimated to be
\beqa \label{eq:Fratio}
\frac{F(1.0833\micron)}{F({\rm Pa}\,\gamma)}
&~\approx~&
\left[\frac{j(1.0833\micron)}{j({\rm Pa}\,\gamma)} ~
\frac{\langle n(\He^+)\rangle}{\langle n(\Ha^+)\rangle}\right] 
\times ~G(\ttot,\xi,\taudzero)
\\ \label{eq:G}
G(\ttot,\xi,\taudzero) &=&
\frac{p_{\rm esc}(\ttot,\xi^{1/3}\taudzero)\exp[-(1-\xi^{1/3})\taudzero]}
{\fesc(\taudzero)}
\eeqa
where $\langle n(\He^+)\rangle$ and $\langle n(\Ha^+)\rangle$ are
electron-density-weighted averages, and $\fesc(\tau)$ is given by
Equation (\ref{eq:F(taud)}).  The quantity in square brackets in
(\ref{eq:Fratio}) is the intrinsic $F(1.0833\micron)/F({\rm
  Pa}\,\gamma)$ ratio; $G$ is the suppresion due to resonant trapping.
The factor $\exp[-(1-\xi^{1/3})\taudzero]$ in Equation (\ref{eq:G}) is
an estimate for the attenuation of the $1.0833\micron$
photons as they traverse the outer zone of the \ion{H}{2} region where
He is neutral.
\begin{figure}
\begin{center}
\includegraphics[angle=0,width=\fwidthb,
                clip=true,trim=0.5cm 5.0cm 0.0cm 4.5cm]
{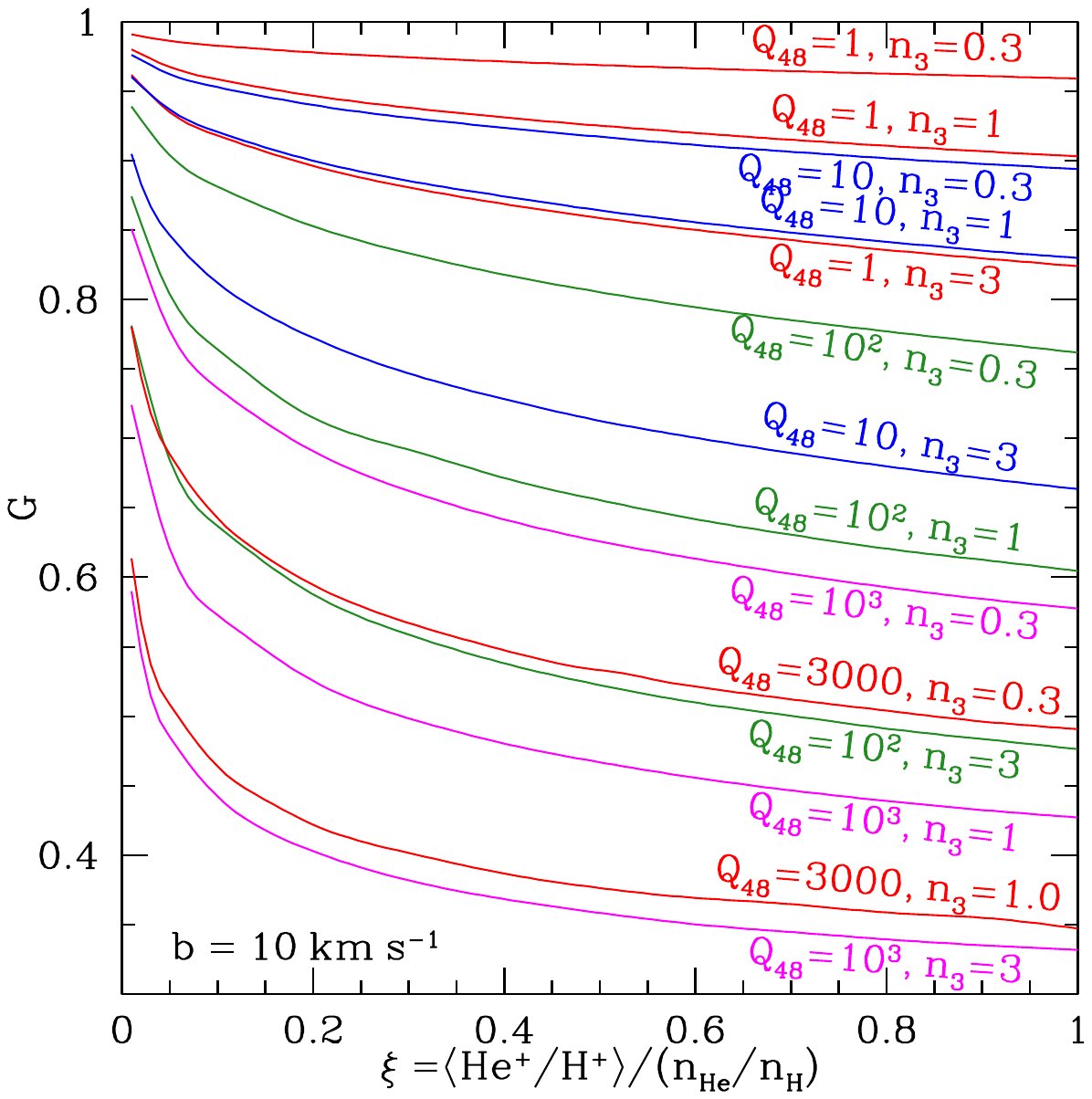}
\includegraphics[angle=0,width=\fwidthb,
                clip=true,trim=0.5cm 5.0cm 0.0cm 4.5cm]
{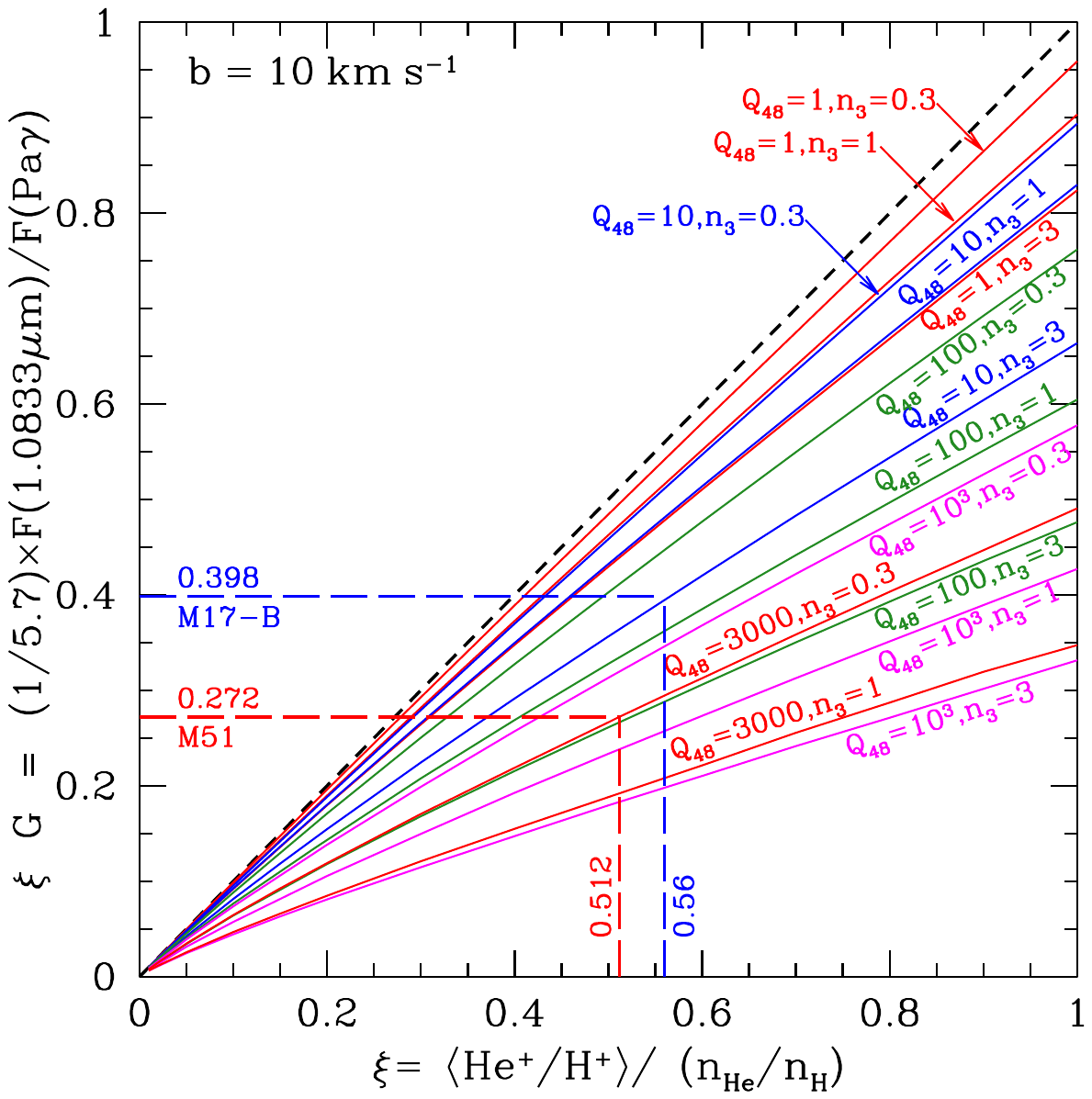}
\caption{\label{fig:g}\footnotesize \newtext{ (a) function $G(\xi)$
    showing extra attenuation of $1.0833\micron$ relative to
    Pa\,$\gamma$ due to resonant trapping, for selected values of
    $Q_{48}$ and $n_3$ (see text).  (b) $\xi G $ as a function of
    $\xi$.  Dashed lines show how $\xi$ (the actual He$^+$/H$^+$
    ratio) can be determined from the observed
    $F(1.0833\micron)/F({\rm Pa}\,\gamma)$ flux ratios for M17-B and
    M51\,NE-Strip\,\ion{H}{2}.}  }
\end{center}
\end{figure}

The function $G(\xi)$ is shown in Figure \ref{fig:g}a, for selected
values of $(Q_{48},n_3)$.  For large values of $Q_{48}$ and $n_3$ (and
$\xi\gtsim 0.1$) $G$ can be significantly smaller than $1$: the
observed flux ratio is significantly smaller than the intrinsic flux
ratio.

An \ion{H}{2} region has $\langle\He^+/\Ha^+\rangle=(n_{\rm
  He}/\nH)\,\xi$.  For case B recombination at $T_e=9000\K$ and
$n_e\approx10^3\cm^{-3}$ we have $4\pi
j(1.0833\micron)=7.065\times10^{-25}\erg\cm^3\s^{-3}$
\citep{DelZanna+Storey_2022} and $4\pi j({\rm
  Pa}\,\gamma)=1.245\times10^{-26}\erg\cm^3\s^{-1}$
\citep{Storey+Hummer_1995}.  Thus, for He abundance $n_\He/\nH=0.10$
we expect
\beq
\frac{1}{5.7} \frac{F(1.0833\micron)}{F({\rm Pa}\,\gamma)}
\approx ~\xi~ G(\ttot,\xi,\taudzero)
~~~.
\eeq
Figure \ref{fig:g}b shows $\xi G = [F(1.0833\micron)/F({\rm
    Pa}\,\gamma)]/5.7$ as a function of $\xi$, the fraction of the He
in the \ion{H}{2} region that is ionized, for selected values of
$(Q_{48},n_3)$.


Figure \ref{fig:g}b can be used to determine the actual value of
$\He^+/\Ha^+$ from the observed ratio
$F(1.0833\micron)/F({\rm Pa}\,\gamma)$: the horizontal red dashed line in
Figure \ref{fig:g}b is the observed value of
$(1/5.7)(F(1.0833\micron)/F({\rm Pa}\,\gamma)=0.272$ for
NE-Strip\,\ion{H}{2} in the galaxy M51
\citep{Draine+Sandstrom+Dale+etal_2025}.  If it is assumed that there
is no differential extinction between $1.0833\micron$ and
Pa\,$\gamma\,1.0941\micron$, we would estimate
$\xi=(\He^+/\Ha^+)/(n_\He/\nH) \approx 0.27$.  However,
NE-Strip\,\ion{H}{2} appears to be characterized by
$Q_{48}=3000,n_3=0.3$; for this case we see that the observed flux
ratio corresponds to $\xi\approx0.51$, 
and $\He^+/\Ha^+\approx (n_\He/\nH) \,\xi\approx 0.051$
-- correcting for the effects of radiative trapping raises the
estimate for $\He^+/\Ha^+$ from $\sim$0.027 to $\sim$0.051.


Similarly, the dashed blue line in Figure \ref{fig:g}b is for the observed
fluxes from M17-B (see Section \ref{subsec:m17b}).  For this case
($Q_{48}\approx 10$, $n_3\approx3$) we estimate $\xi\approx 0.56$
-- correcting for the effects of radiative trapping raises the
estimate for $\He^+/\Ha^+$ from $\sim$0.040 to $\sim$0.056.


\medskip
\section{\label{sec:discuss} Discussion}


The importance of resonant scattering of \ion{H}{1}\,Ly$\alpha$ in
\ion{H}{2} regions has long been recognized, but the effects cannot be
directly observed because Ly$\alpha$ photons escaping from the
\ion{H}{2} region are strongly scattered by \ion{H}{1} outside the
\ion{H}{2} region, and absorbed by interstellar dust.  Thus, it has
not been possible to test predictions for the spectrum of Ly$\alpha$
photons emerging from \ion{H}{2} regions.


The \ion{He}{1}\,$1.0833\micron$ triplet, however, does allow us to
test our theory of resonant scattering.  For bright \ion{H}{2} regions
ionized by early-type O stars, \ion{He}{1}\,$1.0833\micron$ emission
can be strong, and the population of metastable
He$^0$\,\twotripletSone\ is large enough for the optical depth $\ttot$
for resonant scattering of \ion{He}{1}\,$1.0833\micron$ photons within
the \ion{H}{2} region to be large.  Because the population of
metastable He$^0$\,\twotripletSone\ is negligible outside the
\ion{H}{2} region \citep{Indriolo+Hobbs+Hinkle+McCall_2009}, the
\ion{He}{1}\,$1.0833\micron$ photons will travel unimpeded (except by
dust) after escaping the \ion{H}{2} region, allowing remote
observations to test theoretical predictions for the spectrum.


The observable effects of resonant scattering of
\ion{He}{1}\,$1.0833\micron$ will be most conspicuous in high surface
brightness \ion{H}{2} regions with large $\EM_R$.  We consider three
illustrative examples.  }

\begin{figure}
\begin{center}
\includegraphics[angle=0,width=\fwidthb,
                 clip=true,trim=0.5cm 5.0cm 0.0cm 4.5cm]
{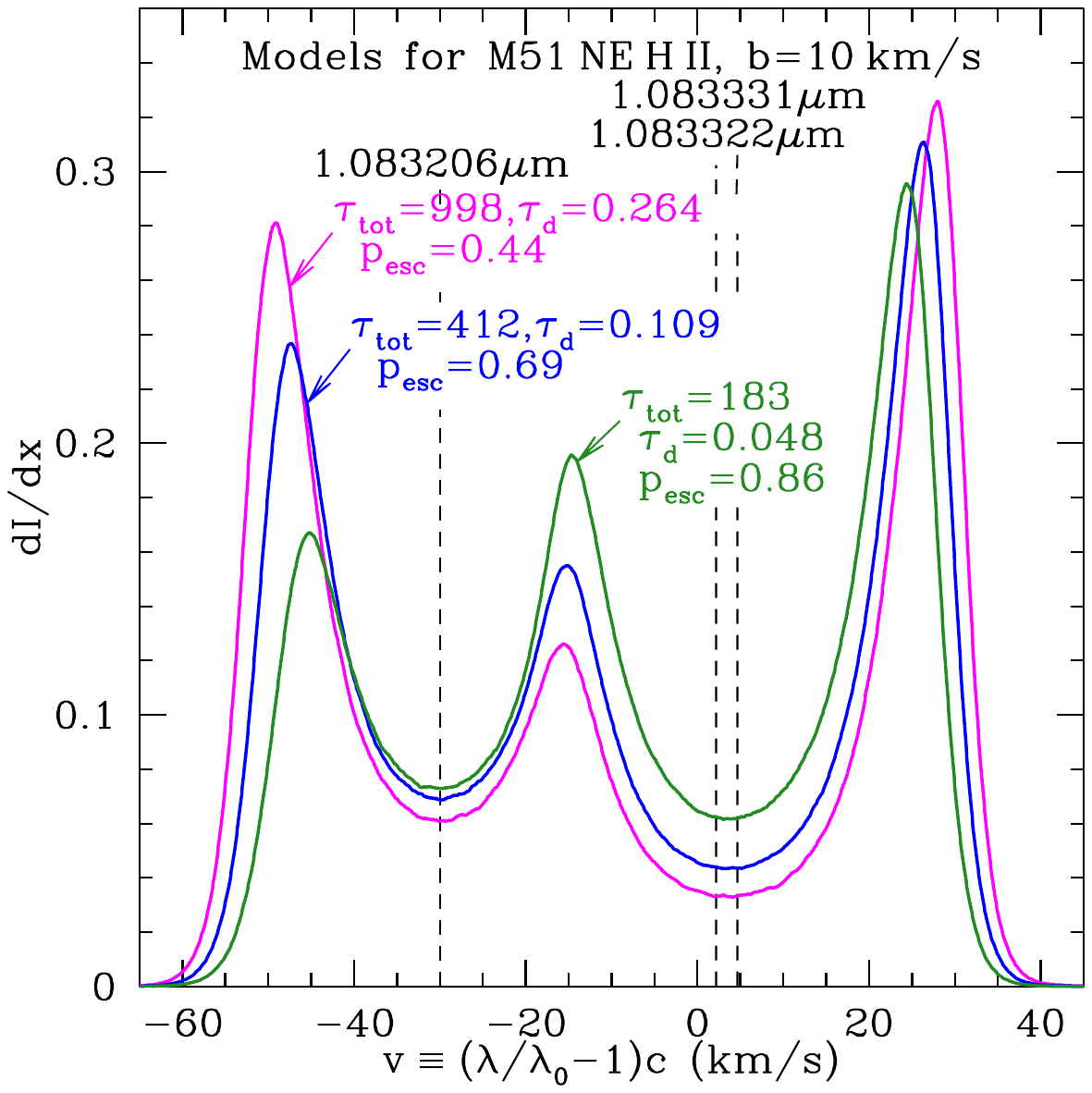}
\caption{\label{fig:m51}\footnotesize \newtext{Models for \ion{He}{1}
    $1.0833\micron$ emission from NE-Strip\,\ion{H}{2} in M51 coming
    from 1, 10, or 100 \ion{H}{2} regions (see text).  The model with
    1 giant \ion{H}{2} region (magenta, $\ttot=998$) has
    $\He^+/\Ha^+\approx 0.057$, the model with 10 giant \ion{H}{2}
    regions (blue, $\ttot=412$) has $\He^+/\Ha^+\approx 0.040$, and
    the model with 100 \ion{H}{2} regions (green, $\ttot=183$) has
    $\He^+/\Ha^+\approx0.035$.  Spectra of escaping photons are
    normalized to $\int (dI/dx) dx=1$.
  }}
\end{center}
\end{figure}

\subsection{Example: Giant \ion{H}{2} Regions in M51}


\omittext{Star-forming galaxies often contain giant \ion{H}{2} regions
  powered by hundreds of O stars.  For example, the}The star-forming
region ``NE-Strip \ion{H}{2}'' in M51 is estimated to be powered by
stars with \omittext{$\dot{N}_{\rm Lyc}\approx
  3\times10^{51}\s^{-1}$,}\newtext{$Q_{48}\approx3\times10^3$}
\citep{Draine+Sandstrom+Dale+etal_2025}.
\citet{Croxall+Pogge+Berg+etal_2015} estimated $n_e\approx
300\pm50\cm^{-3}$ for this region (``NGC5194+91.0+69.0'').
\omittext{If $n_e\approx300\cm^{-3}$, then a single spherical \ion{H}{2} region would have $\EM_R\approx 3\times10^{24}\cm^{-5}$, and Equations (1,12) then
give $\ttot\approx 430(10\kms/b)$.}
\newtext{However, we don't know if the observed emission is dominated
  by a single giant \ion{H}{2} region, or is due to many smaller ones.
  We consider three possibilities:
\begin{enumerate}
\item A single giant \ion{H}{2} region with
$Q_{48}=3000$, $\ttot\approx10^3$, and $\xi=0.57$.
\item Ten giant \ion{H}{2} regions, each with $Q_{48}=300$,
  $\ttot\approx410$, and $\xi=0.40$.
\item 100 smaller \ion{H}{2} regions, each with $Q_{48}=30$,
  $\ttot\approx180$, and $\xi=0.34$.
\end{enumerate}
For all three cases we assume
$n_3\approx 0.3$, and $b\approx10\kms$.


Figure \ref{fig:m51} shows calculated spectra for the three cases.
As the number of individual \ion{H}{2} regions increases,
the central peak increases, and the blue peak decreases.
All cases shown are consistent with the observed $F(1.0833\micron)/F({\rm
    Pa}\,\gamma)\approx 1.7$, but $\He^+/\Ha^+$ varies from $0.057$ to
$0.034$.
If $\xi\approx 0.57$, the stellar population
  should have $Q_1/Q_0\approx 0.088$, corresponding to spectral type
  intermediate between O8.5V and O9V
  \citep{Martins+Schaerer+Hillier_2005}.}


\omittext{Alternatively, the emission may originate from a number of
  unresolved smaller \ion{H}{2} regions, each with lower $\EM_R$ and
  $\ttot$.  If, for examples, we suppose that the emission comes from
  $\sim$30 distinction \ion{H}{2} regions, each with $\dot{N}_{\rm
    Lyc}\approx10^{50}\s^{-1}$, then the emitted spectrum might
  instead be characterized by $\ttot\approx 140(10\kms/b)$.}


\omittext{Figure \ref{fig:m51} shows the triple-peaked
  \ion{He}{1}\,$1.0833\micron$ line profile expected for $b=10\kms$
  and $\ttot=100$, 300, and 500. While the emission profile does
  depend on $\ttot$, in all cases the emission is dominated by two
  peaks, separated by $\sim$75$\kms$, with the red-shifted peak
  slightly stronger than the blue-shifted peak.  There is also a
  weaker central peak near $-16\kms$.  The relative strength of the
  central peak weakens with increasing $\ttot$.

Detection of a line profile resembling the $\ttot=300$ or 500 profiles
in Figure \ref{fig:m51} -- with relatively weak central emission --
would confirm that NE-Strip \ion{H}{2} is dominated by a single giant
\ion{H}{2} region, with emitting gas characterized by a column density
$N(2\,^3{\rm S}_1)\gtsim 10^{15}\cm^{-2}$, and $\ttot \gtsim 400$ (see
Equation 12).

Alternatively, if NE-Strip \ion{H}{2} consistes of a large number of
unresolved smaller \ion{H}{2} regions, the central peak at $-16\kms$
will be stronger than the blue-shifted peak at $\sim$$-45\kms$.  If,
as would be expected, the different \ion{H}{2} have radial velocities
differing by $\ltsim 10\kms$, the three peaks will each be broadened
by the velocity differences between the unresolved \ion{H}{2}
regions.}



$1-1.2\micron$ spectroscopy of this region with the Keck NIRSPEC
spectrograph \citep[$R=25000$;][]{McLean+Becklin+Bendiksen+etal_1998}
or the Calar Alto 3.5m CARMENES spectrograph
\citep[$R=80000$;][]{Quirrenbach+Amado+Caballero+etal_2014} would be
able to test the predictions in Figure \ref{fig:m51}, and distinguish
between the predictions for one giant \ion{H}{2} region, versus
\omittext{tens of}\newtext{many} smaller regions. \newtext{The
  differing estimates of $\He^+/\Ha^+$ could be directly tested using
  radio recombination lines.}

\begin{figure}
\begin{center}
\includegraphics[angle=0,width=\fwidthb,
                 clip=true,trim=0.5cm 5.0cm 0.0cm 4.5cm]
{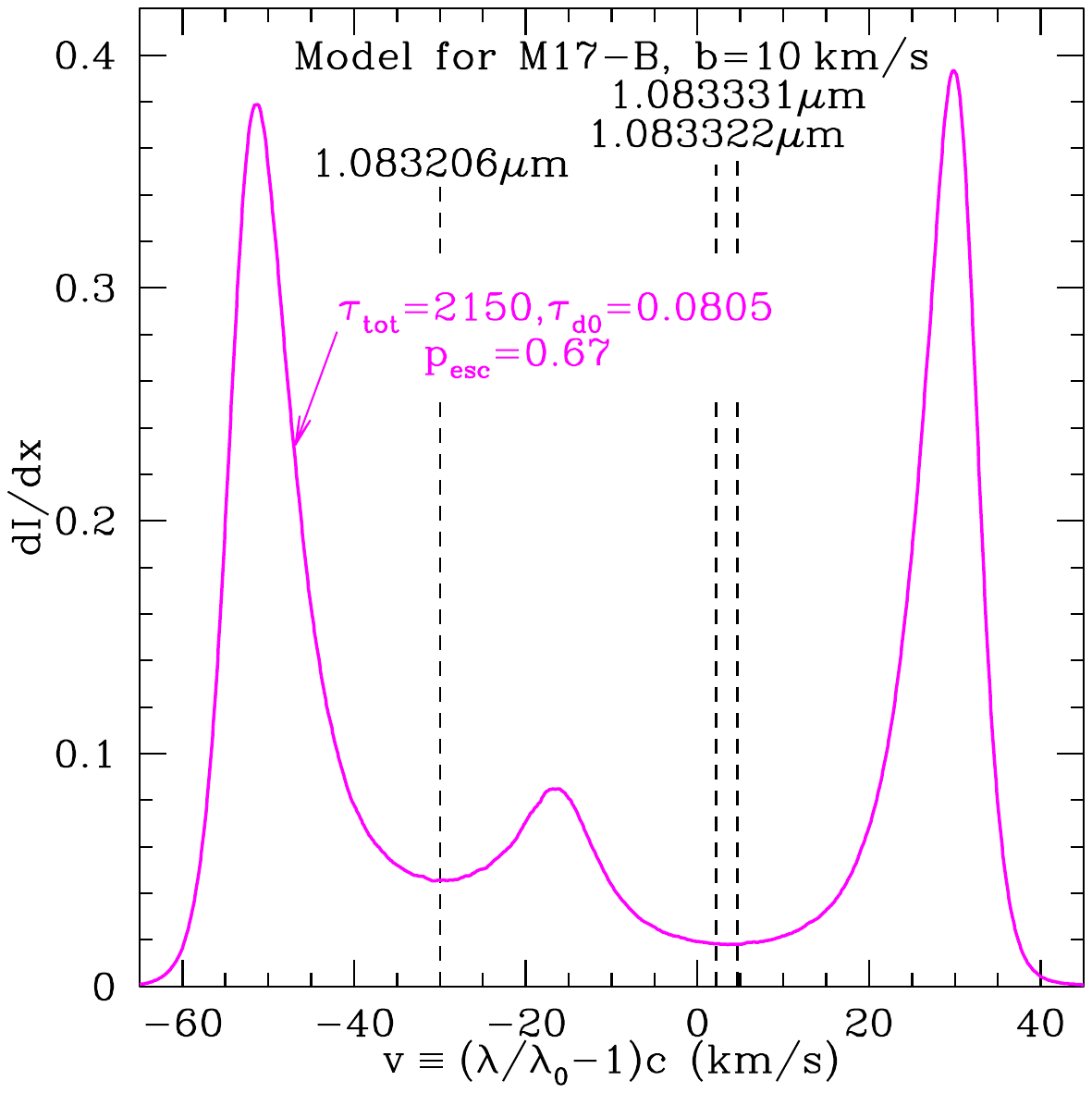}
\caption{\label{fig:m17b}\footnotesize Predicted line profile for
  M17-B (see text) for $b=10\kms$, and $\ttot=2000$
  (see text).  Two strong peaks are expected, at $v\approx-52\kms$ and
  $+30\kms$, and a weaker central peak at $v\approx-16\kms$.
  }
\end{center}
\end{figure}

\subsection{\label{subsec:m17b} Example: M17-B}





%
%


\citet{Boersma+Allamandola+Esposito+etal_2023} obtained JWST NIRSpec
and MIRI-MRS spectra of position M17-B in the M17 \ion{H}{2} region.
\omittext{The M17-B spectrum was modeled with $\EM\approx1.2\times10^{25}\cm^{-5}$, $\He^+/\Ha^+\approx
  0.038$, and $A_V=10.5$\,mag \citep{Draine+Sandstrom+Dale+etal_2025}.

Estimating the center-to-edge $\EM_R\approx
  8\times10^{24}\cm^{-5}$ for a spherical model for the \ion{H}{2}
  region, we estimate center-to-edge $N_R(2\,^3{\rm S}_1)\approx
  6\times10^{14}\cm^{-2}$, and $\ttot\approx 1500$ for $b\approx
  10\kms$.  Figure \ref{fig:m17b} shows spectra calculated for
  $\ttot=500$, $1000$, and $2000$.  The two prominent peaks are
  separated by $\sim$$80\kms$.  The central peak at $-16\kms$ is
  expected to be weak (see Figure \ref{fig:m17b}).  Because the
  strength of the central peak is sensitive to $\ttot$, measuring the
  strength of this peak relative to the stronger outer peaks will
  constrain $\ttot$.}
\newtext{M17-B coincides with the radio continuum peak
  \citep{Matthews+Harten+Goss_1979,Akabane+Sofue+Hirabayashi+Inoue_1989},
  with $\EM\equiv\int n_e
  n(\Ha^+)ds =1.7\times10^{25}\cm^{-5}$ and $n_e\approx3500\cm^{-3}$
  \citep{Matthews+Harten+Goss_1979}.  We note the consistency
  between $\EM\approx1.7\times10^{25}\cm^{-5}$ from the radio
  observations and $\EM\approx1.2\times10^{25}\cm^{-5}$ from modeling
  the recombination lines observed by NIRSpec and MIRI-MRS through
  dust with $A_V\approx10.5\,$mag \citep{Draine+Sandstrom+Dale+etal_2025}.
  
  Here we suppose that the M17-B peak corresponds to an \ion{H}{2}
  region with $Q_{48}=7.4$ \citep[10\% of the estimated ionization
    rate for the entirety of M17;][]{Binder+Povich_2018}.  With
  $n_e\approx3500\cm^{-3}$, this would have peak $\EM=
  2\,\EM_R=2\times10^{25}\cm^{-5}$, consistent with the radio
  determination.
  
  The observed spectrum \citep{Boersma+Allamandola+Esposito+etal_2023} has
  $F(1.0833\micron)/F({\rm Pa}\,\gamma)=2.27$; the inferred He ionization 
  (see Figure \ref{fig:g}b) is
\beq
\frac{\He^+}{\Ha^+} = 0.10~\xi \approx \frac{F(1.0833\micron)/F({\rm Pa}\,\gamma)}{57~G}
= 0.056
\eeq
47\% larger than the value 0.038 obtained without allowing for
the effects of resonant trapping \citep{Draine+Sandstrom+Dale+etal_2025}. 

Figure \ref{fig:m17b} shows a Monte-Carlo calculated spectrum with 
radiative trapping and extinction by dust included, for $b=10\kms$.
Spectroscopy of the \ion{He}{1}\,$1.0833\micron$ \omittext{feature}
\newtext{profile} in M17-B with CARMENES, NIRSPEC, or the VLT
X-shooter \citep[$R=11300$;][]{Vernet+Dekker+DOdorico+etal_2011} can
test whether the \ion{He}{1}\,$1.0833\micron$ emission from M17-B
agrees with the predicted spectra in Figure \ref{fig:m17b}.
}

\begin{figure}
\begin{center}
\includegraphics[angle=0,width=\fwidthb,
                 clip=true,trim=0.5cm 5.0cm 0.0cm 4.5cm]
{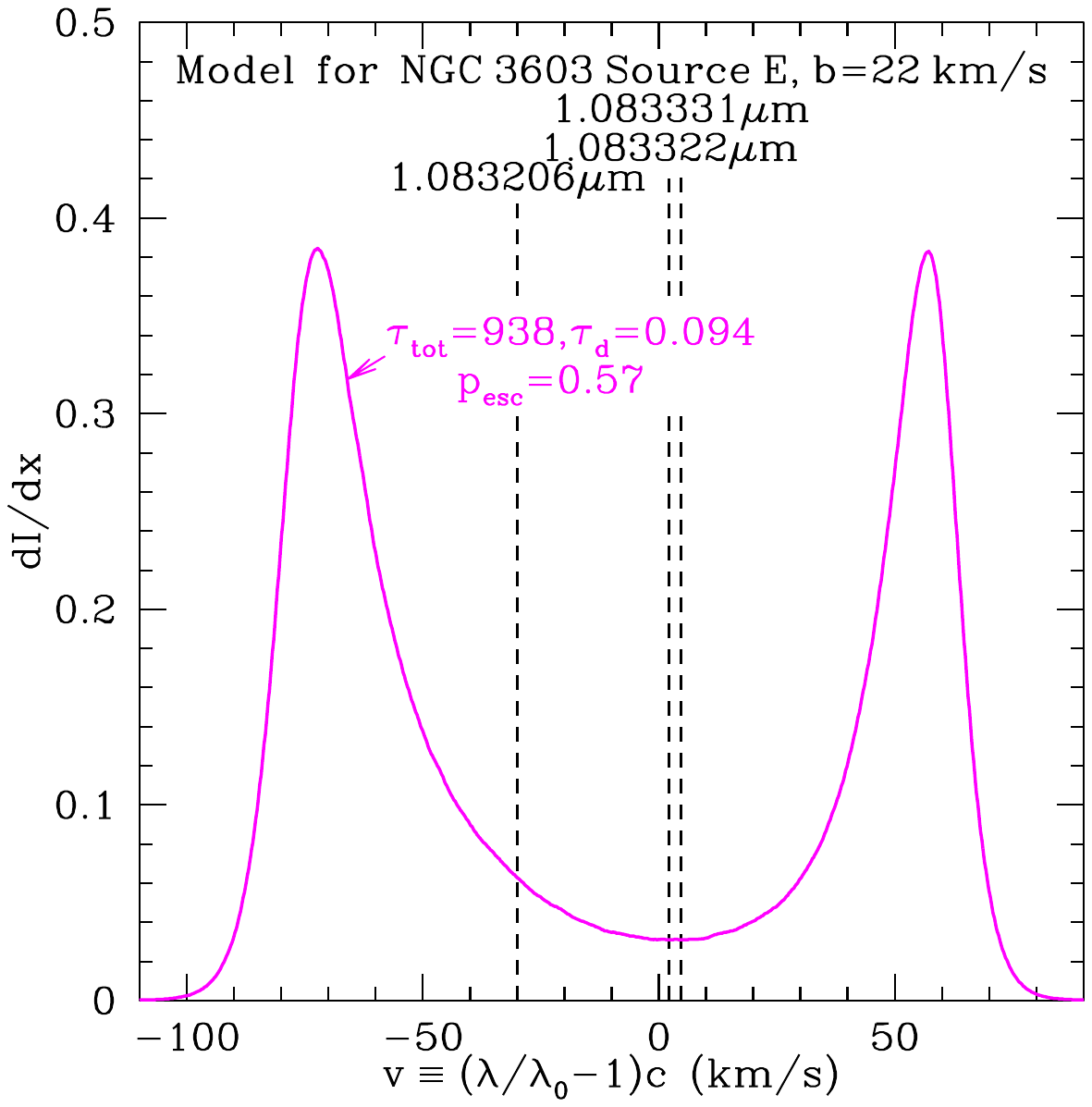}
\caption{\label{fig:n3603}\footnotesize Predicted line profile for NGC
  3603 Source E for $b=22\kms$, and $\ttot=1300$, with dust absorption
  (see text).  Two strong peaks are expected, at $v\approx-80\kms$ and
  $+60\kms$.
  }
\end{center}
\end{figure}

\subsection{Example: NGC\,3603}

The giant \ion{H}{2} region NGC 3603 has a number of locations with
very high \EM.  Radio continuum observations of Source E with
$7\arcsec$ resolution find $\EM\approx 1.4\times10^{25}\cm^{-5}$
\citep{DePree+Nysewander+Goss_1999}; we estimate the
center-to-edge $\EM_R\approx 1.0\times10^{25}\cm^{-5}$ for Source E.

The observed radio recombination lines imply electron temperature
$T_e\ltsim 9000\K$ \citep{DePree+Nysewander+Goss_1999}.  The
He90$\alpha$ line profile is consistent with $b\approx22\kms$,
requiring that the line width be dominated by a combination of
turbulence and fluid flow, possibly expansion.  The present
calculations do not include systematic flow: we assume that the
observed $b=22\kms$ is due to a combination of thermal broadening and
microturbulence.
  
The He90$\alpha$/H90$\alpha$ line ratio is consistent with the He
being fully ionized ($\He^+/\Ha^+\approx 0.10$), so we take
$\xi\approx1$.  From Equation (\ref{eq:ttot_vs_Q48n3}) we estimate
$\ttot\approx 940$.  Figure \ref{fig:n3603} shows the predicted line
profile for $b=22\kms$ and $\ttot=940$ \newtext{including dust.}  The
  predicted spectrum has $\FWHM> 150\kms$.  We predict that the
  \ion{He}{1} $1.0833\micron$ emission will appear as two strong
  peaks, at $v\approx -80\kms$ and $+60\kms$.

Spectral resolution $R\equiv\lambda/\Delta\lambda \gtsim 2\times10^4$
would be required to measure the predicted $\sim$$20\kms$ widths of
each of the two predicted peaks, which would be possible with
instruments such as the WINERED spectrograph
\citep{Otsubo+Ikeda+Kobayashi+etal_2016} on Magellan.  However,
$R=11300$ ($\Delta v=27\kms$) spectroscopy with the VLT X-shooter
\citep{Vernet+Dekker+DOdorico+etal_2011} would be sufficient to
measure the predicted $\sim$$140\kms$ splitting of the two peaks in
the profile, and confirm the deep central minimum predicted in Figure
\ref{fig:n3603}.


\newtext{
\subsection{Other Processes Affecting \ion{He}{1}\,$1.0833\micron$ Emission}

Above we have been assuming that \ion{He}{1}\,$1.0833\micron$ photons
are injected as a result of standard ``case B'' recombination of
$\He^+$ with thermal electrons.  However, the rate of injection of
$1.0833\micron$ photons can be larger due to two processes:
\begin{enumerate}
\item
He atoms that are in the \twotripletSone\ metastable level
can undergo inelastic electron collisions that excite the
\twotripletPJ\ level, followed by emission of an additional
$1.0833\micron$ photon.  The fractional increase in $1.0833\micron$
emission due to collisional excitation of \twotripletPJ\ is
\beq
\frac{n_e k_e}{A_{\rm ms}+n_e k_d}
~~~,
\eeq
where $k_e$ is the rate coefficient for $2^3{\rm S}_1 \rightarrow
2^3{\rm P}$ and $k_d\approx 3.3\times10^{-8}\cm^3\s^{-1}$ is the rate
coefficient for collisional transitions from \twotripletSone\ to
singlet states.  However, $k_e/k_d \ltsim 0.014$ for $T\ltsim 10^4\K$
\citep[using rates from][]{Bray+Burgess+Fursa+Tully_2000} so this
process can be neglected.


%

\item When $N(2\,^3{\rm S}_1)$ is large, radiative trapping can ``convert''
$3889.7\Angstrom$ photons into $4.296\micron$, $7067.1\Angstrom$, and
$1.0833\micron$ photons (see Figure \ref{fig:He levels}), as
originally pointed out by \citet{Pottasch_1962}.  However, this can
increase the emission of $1.0833\micron$ photons by at most 21\% for
$\ttot\gtsim 10^3$, so this is not a major correction.
\end{enumerate}
}

\subsection{\ion{He}{1} Line Ratios}

Resonant trapping of \ion{He}{1} $1.0833\micron$ and other permitted
transitions to \twotripletSone\ (e.g., \twotripletSone\ --
\threetripletP\,$3890\Angstrom$ and \twotripletSone\ --
\fourtripletP\,$3189\Angstrom$) can suppress the power in these lines,
and enhance the emission in other lines (e.g., \threetripletSone\ --
\twotripletP\,$7067\Angstrom$), as originally pointed out by
\citet{Pottasch_1962}.

A number of authors have estimated the effects of radiative transfer
on observed line \ion{He}{1} line ratios
\citep[e.g.,][]{Robbins_1968b, Osterbrock_1989,
  Benjamin+Skillman+Smits_2002, Blagrave+Martin+Rubin+etal_2007},
using different estimates for the ``escape probability'' $\beta$ for
\ion{He}{1}\,$1.0833\micron$ photons.  The present Monte-Carlo study
provides improved estimates for $\beta$ from a spherical nebula, with
values of $\beta$ that can be significantly below previous estimates
(see Figure \ref{fig:beta}). A future paper (B.T.~Draine 2026, in
preparation) will use the escape probabilities obtained here to
recalculate the effects of \omittext{radiative
  transfer}\newtext{resonant scattering} on \ion{He}{1} line ratios.

\section{\label{sec:summary} Summary}

\begin{enumerate}

\item Monte-Carlo radiative transfer calculations have been carried
  out for resonant scattering by metastable He$^0$\,\twotripletSone\ in
  \newtext{dustless} \ion{H}{2} regions, using exact partial
  redistribution and Voigt line profiles.  The mean number of
  scatterings $\langle N_{\rm sca}\rangle$ and the mean pathlength
  $\langle L_{\rm path}\rangle$ traveled for \ion{He}{1}
  $1.0833\micron$ triplet photons were determined as functions of
  \ion{He}{1}\,$1.0833\micron$ optical depth $\ttot$, for different
  values of $b$.  Simple fitting formulae for $\langle N_{\rm
    sca}\rangle$ and $\langle L_{\rm path}\rangle$ are given.

\item Resonant scattering by metastable He$^0$\,\twotripletSone\ can
  produce unusual \newtext{multi-peaked} line profiles for the
  \ion{He}{1} $1.0833\micron$ triplet emission from \ion{H}{2}
  regions.

\item For normal \ion{H}{2} region conditions ($b\approx10\kms$,
  $10^2\ltsim\ttot\ltsim 10^3$), the overall
  \ion{He}{1}\,$1.0833\micron$ emission can be blue-shifted by up to
  $\sim$$14\kms$.

\newtext{
\item For \ion{H}{2} regions with $Q_{48}n_3\gtsim 10$, dust can
  significantly reduce the emission in \ion{He}{1}\,$1.0833\micron$,
  and influence the shape of the emission profile.  Estimates of
  He$^+$/H$^+$ using
  \ion{He}{1}\,$1.0833\micron$/\ion{H}{1}\,Pa$\gamma$ should allow for
  the resonant-scattering-enhanced attenuation of
  \ion{He}{1}\,$1.0833\micron$ by dust.  Monte-Carlo calculations have
  been carried out to explore the effects of dust, for dust abundances
  appropriate to star-forming galaxies with near-solar metallicities.

\item For Milky Way dust abundances, we estimate the effects of
  resonant trapping on the $\He\,1.0833\micron/{\rm
    H\,Pa}\,\gamma\,1.0941\micron$ line ratio.  If $Q_{48}$ and $n_3$
  can be estimated, Figure \ref{fig:g}b can be used to estimate
  $\He^+/\Ha^+$.

\item In low metallicity systems, dust attenuation will be less
  important, but resonant trapping will continue to produce the
  unusual multipeaked line profiles found here.}

\item We predict \ion{He}{1}\,$1.0833\micron$ line profiles for the
  Galactic \ion{H}{2} regions M17-B and NGC 3603, and for the
  star-forming region ``NE-Strip \ion{H}{2}'' in the Whirlpool galaxy
  M51.  The profiles typically consist of two strong peaks, separated
  by $\sim$75--140$\kms$, with the centroid of the
  \ion{He}{1}\,$1.0833\micron$ emission blue-shifted by $\sim$$13\kms$
  relative to other lines.  For modest $\ttot$ a central peak at
  $-18\kms$ is also present.  The relative strength of the
  central peak decreases as $\ttot$ is increased.  The splitting, blue
  shift, and central peak could be confirmed with $R\gtsim 10^4$
  spectroscopy.
\end{enumerate}

\begin{acknowledgements}

I am grateful to Kwang-il Seon for very helpful communications
regarding Ly\,$\alpha$ scattering\newtext{, and to the anonymous
  referee for comments that led to improvement of the manuscript}.  I
thank Giulio Del Zanna, Jim Gunn, and Dina Gutkowicz-Krusin for
helpful discussions\newtext{, and Robert Lupton for continued availability of
the SM plotting package}.
\end{acknowledgements}


\begin{appendix}

\section{Monte-Carlo Radiative Transfer}
\subsection{Opacity}

Consider scatterers with lower level $E_\ell$ (e.g.,
the metastable $1s2s\,^3{\rm S}_1$ state) and $N$ excited states, $E_u$,
$u=1-N$, with degeneracies $g_u$, Einstein $A$ coefficients
$A_{u\ell}$, and slightly different wavelengths $\lambda_{u\ell}$.
It is convenient to measure the photon frequency by the dimensionless
Doppler shift $x_j$ relative to the multiplet centroid $\lambda_c$,
\beq
x = \frac{\lambda_c/\lambda-1}{b/c}
~~~.
\eeq
The dimensionless Doppler shift of the photon relative to the local
$\lambda_{u\ell}$ line is $x+\Delta_u$ where
\beq \label{eq:Delta_k}
\Delta_u \equiv \left(\frac{\lambda_{u\ell}}{\lambda_c}-1\right)\frac{c}{b}
~~~.
\eeq
The attenuation coefficient 
\beqa
\alpha(x) &~=~& \nH\sigmad + n({\rm He\,2\,^3S_1})
\sum_{u=0}^2 \sigma_{\ell u}(x) 
\\
\label{eq:voigt}
\sigma_{\ell u}(x) &=& 
\frac{g_u A_{u\ell}\lambda_{u\ell}^3}{8\pi^2 b}
\phi(x+\Delta_u,a)
~~~,
\eeqa
\newtext{where $\sigmad$ is the dust absorption cross section
per H nucleon.}
The Voigt line profile $\phi(x,a)$ is obtained by table look-up from
tables generated using approximations from \citet{Armstrong_1967}.

%
%

%
%

\subsection{Injection}

Photons are injected uniformly throughout the spherical volume, with
\newtext{initial} frequencies $x$ drawn from the Voigt line profile
$\phi(x)$, and random directions of propagation $\bnhat$.
\newtext{All results for regions with dust were calculated using at
  least $10^6$ (in most cases $10^7$) injected photons to ensure
  statistical accuracy.}

\subsection{Transport}

Let $s$ measure path length in line-center optical depth units:
\beq
ds = \ttot\frac{dL}{R}
~~~,
\eeq
\newtext{where $L$ is physical path length.}  Let
$t\newtext{=(r/R)\ttot}$ be the line-center optical depth from the
center of the sphere to a location \newtext{at radius $r$}.
\newtext{Let $\taud=\nH\sigmad R$ be the dust absorption optical depth
  from center to edge.}
 
Let $t_j$ be the location of a photon with Doppler shift $x_j$,
traveling in direction $\bnhat_j$.  
Define the direction cosine
\beq
\mu_j \equiv \bnhat_j\cdot\brhat_j
~~~,
\eeq
where $\brhat_j$ is the radial direction at the photon's location.  To
escape, the photon must traverse a path length
\beq
\s_{esc} =
\left\{
\left[\ttot^2-t_j^2(1-\mu_j^2)\right]^{1/2} - t_j\mu_j
\right\}
~~~.
\eeq
The probability of escape for this photon is
\beqa
P_{{\rm esc}} &~=~& \exp\left[-\tau(s_{\rm esc})\right]
\\
\tau(s) &=& \int_0^{s} \alpha(x_j,s^\prime)\,\frac{R}{\ttot} ds^\prime
~~~.
\eeqa
The probability of scattering or absorption in an interval $[s,s+ds]$ is
\beq
dP = \exp\left[-\tau(s)\right]~ d\tau
~~~.
\eeq
A random number $\prand_1$ is drawn\footnote{%
\newtext{We employ the Mersenne Twister random number generator
\citep{Matsumoto+Nishimura_1998}, translated into Fortran by Tsuyoshi Tada.}}
from a distribution uniform on $[0,1]$.  The photon escapes if
$\prand_1>1-P_{\rm esc}$; if $\prand_1<1-P_{\rm esc}$ there is an
event (either scattering or absorption) after traversing optical depth
\beq
\tau_j = -\ln(1-\prand_1)
~~~.
\eeq
If we take $\alpha(x_j)$ to be constant along the path, then
\beq
s_j = \frac{-\ln(1-\prand_1)}{\alpha(x_j)}
~~~.
\eeq
After traveling a pathlength $s_j$, the event \newtext{(scattering or
absorption)} will be at
\beq
\label{eq:tnew}
t_{j+1} = \left[ t_j^2 + 2\mu_j t_j s_j + s_j^2\right]^{1/2}
~~~,
\eeq
with new direction cosine
$\tilde{\mu}_{j+1}\equiv\bnhat_j\cdot\brhat_{j+1}$:
%
% ...
%
\beqa 
\tilde{\mu}_{j+1} &~=~& ~~
\left[1-\left(\frac{t_j}{t_{j+1}}\right)^2(1-\mu_j^2)\right]^{1/2}
\hspace*{10mm}{\rm if~}s>-\mu_jt_j
\\
&=&
-\left[1-\left(\frac{t_j}{t_{j+1}}\right)^2(1-\mu_j^2)\right]^{1/2}
\hspace*{10mm}{\rm if~}s<-\mu_jt_j
\eeqa
\emph{prior} to scattering (or absorption).

The probability that the event is resonant scattering by transition
$\ell\rightarrow u$ is
\beq
P_{\ell u} = \frac{n_\ell \sigma_{\ell u}(x_j)}{\alpha(x_j)}
%
~~~.
\eeq
%
%
%
%
%
A random number $\prand_2\in[0,1]$ is drawn to determine whether the
event is resonant scattering by transition $\ell\rightarrow u$.
\newtext{If $\prand_2>\sum_{u=0}^2 P_{\ell u}$, the photon is absorbed
  by dust.}

  
\subsection{Direction After Scattering}

Suppose that scattering takes place at $t_{j+1}$, with scattering
angle $\gamma_{j+1}$.  The new direction of propagation $\bnhat_{j+1}$
will have direction cosine
\beq
\mu_{j+1}\equiv\bnhat_{j+1}\cdot\brhat_{j+1} = 
\tilde{\mu}_{j+1}\,\cos\gamma_{j+1}+
(1-\tilde{\mu}_{j+1}^2)^{1/2}\sin\gamma_{j+1}\,\cos\phi_{j+1}
\eeq
and new dimensionless Doppler shift $x_{j+1}$.
For unpolarized light, $\phi_{j+1}$ is uniformly distributed in $[0,2\pi]$:
\beq \label{eq:dPdphi}
\frac{dP_\phi}{d\phi_{j+1}} = \frac{1}{2\pi} ~~~,~~~ \phi_{j+1}\in[0,2\pi]
~~~.
\eeq
and $\cos\gamma_{j+1}$ has probability distribution
\beq \label{eq:dPdgamma}
\frac{dP}{d\cos\gamma} = \frac{3}{8}\left(1+\cos^2\gamma\right) 
~~,~~~\cos\gamma\in[-1,1]
~~~.
\eeq
Equation (\ref{eq:dPdgamma}) applies for both resonant scattering
(e.g., by \ion{He}{1}\,\twotripletSone) and Thomson scattering by free
electrons.  After drawing a random variable $P_\gamma$ uniformly
distributed on $[0,1]$, we obtain $\cos\gamma$ using the result found
by \citet{Seon_2006}:
\beq \label{eq:P(gamma)}
\cos\gamma = \left[2(2P_\gamma-1)+\sqrt{4(2P_\gamma-1)^2+1}\right]^{1/3}
- \left[2(2P_\gamma-1)+\sqrt{4(2P_\gamma-1)^2+1}\right]^{-1/3}
~~~.
\eeq

\subsection{Frequency After Scattering}

Let $\bw_{j+1}$ be the velocity of the
scatterer in the fluid frame, at the location $t_{j+1}$ of the
scattering.  We are concerned only with components of
$\bw_{j+1}$ in the scattering plane.  Define
\beqa
w_{\parallel,j+1} &~\equiv~&\bnhat_j\cdot\bw_{j+1}
\\
w_{\perp,j+1}&\equiv&
[(\bnhat_{j+1}-\cos\gamma_{j+1} \bnhat_j)\cdot\bw_{j+1}]/\sin\gamma_{j+1}
~~~.
\eeqa
Neglecting recoil, the dimensionless frequency after scattering is
\beq \label{eq:xnew}
x_{j+1}=x_j+
w_{\parallel,j+1}
\left(\cos\gamma_{j+1}-1\right)
+
w_{\perp,j+1} \sin\gamma_{j+1}
~~~.
\eeq

We use the ``partial redistribution function'' for a line with finite
intrinsic width \citep{Hummer_1962}.  The exact distribution for
$w_\parallel$ is difficult to invert.  For given $x_j$ and transition
$\ell\rightarrow u$, following \citet{Zheng+Miralda-Escude_2002}, we
use the ``method of rejection''
\citep{Press+Teukolsky+Vetterling+Flannery_1992} to draw
$w_{\parallel,j+1}$ from the exact distribution
\beq \label{eq:dPdu_parallel}
\frac{dP}{dw_{\parallel,j+1}} \propto
\frac{e^{-w_{\parallel,j+1}^2}}{(x_j+\Delta_u-w_{\parallel,j+1})^2+a^2}
~~~.
\eeq
We employ a modification of the ``comparison function'' $g(x)$ used by
\citet{Zheng+Miralda-Escude_2002}: for $x\geq0$:
\beqa
g(x) &~=~& \frac{1}{(x-w_\parallel)^2+a^2} \hspace*{10mm} 0\leq x \leq C_1
\\
&=& \frac{e^{-C_1^2}}{ (x-w_\parallel)^2+a^2} \hspace*{10mm} C_1 < x \leq C_2
\\
&=& \frac{e^{-C_2^2}}{ (x-w_\parallel)^2+a^2} \hspace*{10mm} C_2 < x \leq C_3
\\
&=& \frac{e^{-C_3^2}}{ (x-w_\parallel)^2+a^2} \hspace*{10mm} C_3 < x \leq C_4
\\
&=& \frac{e^{-C_4^2}}{ (x-w_\parallel)^2+a^2} \hspace*{10mm} C_4 < x
~~~.
\eeqa
For $|x|\leq1$ we take $C_1=0.5$, $C_2=0.8$, $C_3=1.2$, and $C_4=2.5$;
for $|x|>1$, $C_1=|x|/2$, $C_2=|x|-0.2$, $C_3=|x|+0.2$, and
$C_4=|x|+1.5$.  This greatly increases the efficiency of the method,
by reducing the number of rejections.

For given $w_{\parallel,j+1}$, the transverse velocity $w_{\perp,j+1}$ is
drawn from the thermal distribution; the probability distribution for
$x_{j+1}$ is gaussian:
%
%
\beq
x_{j+1} = x_j+\left(\cos\gamma_{j+1}-1\right)
              w_{\parallel,j+1}
+ \frac{\sin\gamma_{j+1}}{\sqrt{2}} \,\prand
~~~,
\eeq
where $\prand$ is a Gaussian random variate with zero mean and unit
variance.


%
%
%
%
\end{appendix}

\bibliography{/u/draine/work/libe/btdrefs}
\bibliographystyle{aasjournal}

\end{document}